\documentclass[a4paper,12pt]{article}

\usepackage[usenames]{color}
\usepackage{graphicx}
\usepackage{amsmath}
\usepackage{amssymb}
\usepackage{setspace}
\usepackage{xcolor}

\begin{document}
\begin{titlepage}
\null
\begin{flushright}
June, 2017
\end{flushright}

\vskip 1.8cm
\begin{center}

  {\Large \bf ADHM Construction of (Anti-)Self-dual Instantons\\
\vspace{0.5cm}
in Eight Dimensions}

\vskip 1.8cm
\normalsize

  {\bf Atsushi Nakamula\footnote{nakamula(at)sci.kitasato-u.ac.jp},
Shin Sasaki\footnote{shin-s(at)kitasato-u.ac.jp} and Koki Takesue\footnote{ktakesue(at)sci.kitasato-u.ac.jp}}

\vskip 0.5cm

  { \it
  Department of Physics \\
  Kitasato University \\
  Sagamihara 252-0373, Japan
  }

\vskip 2cm

\begin{abstract}
We study the ADHM construction of
(anti-)self-dual instantons in eight dimensions.
We propose a general scheme to construct the (anti-)self-dual gauge field
 configurations $F \wedge F = \pm *_8 F \wedge F$ whose finite
 topological charges are given by the fourth Chern number.
We show that
our construction reproduces
 the known $\text{SO}(8)$ one-instanton solution.
We also construct multi-instanton solutions of
the 't Hooft and the Jackiw-Nohl-Rebbi (JNR) types
in the dilute instanton gas approximation.
The well-separated configurations of
multi-instantons reproduce the correct topological charges with high accuracy.
We also show that our construction is generalized to
(anti-)self-dual instantons in $4n \ (n=3,4, \ldots)$ dimensions.
\end{abstract}
\end{center}
\end{titlepage}

\newpage
\tableofcontents
\section{Introduction}
It is well known that instantons in gauge theories play important roles
in the study of non-perturbative effects \cite{Belavin:1975fg, 'tHooft:1976fv}.
Instantons in four-dimensional gauge theories with gauge group $G$
are defined by configurations
such that the gauge field strength 2-form $F$ satisfies the
(anti-)self-dual equation $F = \pm*_4 F$.
Here $*_d$ is the Hodge dual operator in $d$-dimensional Euclid space.
Due to the Bianchi identity, instanton solutions in four-dimensional
Yang-Mills theory satisfy the equation of motion.
The instanton solutions are classified by the second Chern number which is
proportional to $\int \! \mathrm{Tr} [F \wedge F]$.
They are characterized by the homotopy group $\pi_3 (G)$.
A salient feature of the
(anti-)self-dual instantons in four dimensions is its
systematic construction of solutions, known as the ADHM construction \cite{Atiyah:1978ri}.
The ADHM construction reveals the K\"ahler quotient structure of
the instanton moduli space and provides the scheme to calculate the
non-perturbative corrections in the path integral.

It is natural to generalize the instantons in four dimensions to higher
and lower dimensions.
In the lower dimensions, the dimensional reduction of the
(anti-)self-dual equation to three dimensions leads to the monopole equations.
The ADHM construction is reduced to the Nahm construction of the monopoles \cite{Nahm:1979yw}.
In two dimensions, the (anti-)self-dual equations provide equations for
the Hitchin system \cite{Hitchin:1986vp}. Further dimensional reductions of the
(anti-)self-dual equation give equations in various integrable systems in
one and two dimensions \cite{Mason:1991rf}. This is known as the Ward's conjecture \cite{Ward:1985gz}.

On the other hand, instantons in dimensions higher than four have been
studied in various contexts.
It is known that there are several kinds of ``instantons'' in higher
dimensions.
A straightforward generalization of the (anti-)self-dual equation $F = \pm *_4 F$
 to $d>4$ dimensions is the linear equation $F_{\mu \nu} = \lambda
 T_{\mu \nu \rho \sigma}
F^{\rho \sigma}, \lambda \not= 0, \ (\mu,\nu,\rho,\sigma = 1, \ldots, d)$ \cite{Corrigan:1982th,Dunajski:2011sx,Devchand:2012xs}.
Here $T_{\mu \nu \rho \sigma}$ is an anti-symmetric constant tensor which
respects subgroups of the $\text{SO}(d)$ Lorentz group.
This equation is called the secular type and
solutions to this equation are
sometimes called secular type instantons.
Note that the secular type instantons satisfy the equation of motion for
Yang-Mills theory but it is not always true that 
Chern numbers associated with the solutions are finite and quantized.
An example of the secular type instanton
is the Fubini-Nicolai instantons \cite{Fubini:1985jm}, also known as octonionic
instantons, defined in eight dimensions.
Other examples are BPS instantons that preserve fractions of
supersymmetry in eight-dimensional super Yang-Mills theory
\cite{Bak:2002aq}.
An ADHM construction of secular type instantons in $4n \ (n=1,2,3,\ldots)$
dimensions has been studied \cite{Corrigan:1984si}.

Among other things, instantons in $4n \ (n=1,2,3,\ldots)$ dimensions provide special
interests. This is because in these dimensions, the (anti-)self-dual equations of the field strengths are
naturally defined.
For example, in eight dimensions $(n=2)$, we can
define the (anti-)self-dual equation $F \wedge F = \pm *_8 F \wedge F$.
We call solutions to this equation
the (anti-)self-dual instantons in eight dimensions.
We expect that configurations which satisfy the (anti-)self-dual equation have
non-zero topological charges given by the fourth Chern number
$k =\mathcal{N}
\int \!
\mathrm{Tr} [F\wedge F \wedge F \wedge F]$,
where $\mathcal{N}$ is a normalization constant.
Since the eight-dimensional (anti-)self-dual
equation is highly non-linear and contains higher derivatives,
only the one-instanton solution is known \cite{Grossman:1984pi, Tchrakian:1984gq}.
This is called the $\text{SO}(8)$ instanton.
Note that the $\text{SO}(8)$ instanton does not satisfy the secular equation in general.

In this paper, we study an ADHM construction of (anti-)self-dual instantons in
eight dimensions.
We will show that there is a general scheme to find the (anti-)self-dual
instanton solutions.
By introducing specific ADHM data which solve ADHM constraints, 
we will explicitly construct gauge
field configurations whose fourth Chern numbers are integers.
This implies that the solutions are characterized by the homotopy group
$\pi_7(G)$.
We will also discuss eight-dimensional higher derivative theories in
which the (anti-)self-dual equation $F \wedge F = \pm *_8 F \wedge F$ becomes relevant.

The organization of this paper is as follows.
In the next section, we study the ADHM construction of
(anti-)self-dual instantons in eight dimensions. This is just an eight-dimensional analogue of the
original ADHM construction of instantons in four dimensions.
We find that there is an extra ADHM constraint in addition to
the original one which is present in four dimensions.
The gauge group and algebraic structures of the solutions are studied in detail.
In section 3,
we see that our construction precisely reproduces the well-known
one-instanton profile of the solution \cite{Grossman:1984pi, Tchrakian:1984gq}.
Furthermore we construct the so-called 't Hooft and the Jackiw-Nohl-Rebbi (JNR) type
multi-instantons.
The ADHM data associated with these solutions solve the ADHM constraints
in the dilute instanton gas limit.
We obtain the correct topological charges in a good accuracy.
In section 4, we discuss eight-dimensional gauge field theories where the (anti-)self-dual
instantons are analyzed.
We observe that the multi-instantons of the 't Hooft type can be
interpreted as D$(-1)$-branes embedded in the D7-branes in the small instanton limit.
Section 5 is devoted to conclusion and discussions.
The ADHM construction of instantons in four dimensions is briefly
discussed in Appendix A.
The Clifford algebras in $4n$ dimensions are shown in Appendix B.
The explicit form of the eight-dimensional ADHM equations for the gauge
group U(8) is found in Appendix C.

\section{ADHM construction in eight dimensions}
In this section, we study the ADHM construction of (anti-)self-dual
instantons in eight-dimensional Euclid space with the flat metric.
The (anti-)self-dual equation is defined by
\begin{align}
&F\wedge F = \pm\ast_8 \left( F\wedge F \right),
\label{eq:self-dual_R8_1}
\end{align}
where the 2-form $F = \frac{1}{2!} F_{\mu \nu} dx^{\mu} \wedge dx^{\nu}$
is the gauge field strength whose component is defined by
\begin{align}
F_{\mu \nu} = \partial_{\mu} A_{\nu} - \partial_{\nu} A_{\mu} +
 [A_{\mu}, A_{\nu}].
\end{align}
The anti-Hermite gauge field $A_{\mu}$ takes value in $\mathcal{G}$.
Here $\mathcal{G}$ is the Lie algebra associated with the non-Abelian
gauge group $G$ and $\mu,\nu,\ldots=1,2,\ldots,8$ are the tensor indices in
the eight-dimensional Euclid space.
The (anti-)self-dual equation
\eqref{eq:self-dual_R8_1} in the component expression is given by
\begin{align}
F_{[\mu\nu}F_{\rho\sigma]} = \pm
 \frac{1}{4!}\varepsilon_{\mu\nu\rho\sigma\alpha\beta\gamma\delta}F_{\alpha\beta}F_{\gamma\delta},
\label{eq:self-dual_R8_2}
\end{align}
where $\varepsilon_{\mu\nu\rho\sigma\alpha\beta\gamma\delta}$ is the
anti-symmetric epsilon symbol in eight dimensions and the bracket $[\mu_1
\mu_2 \cdots \mu_n]$ stands for the anti-symmetrization of indices with
the weight $1/n!$. In the following subsections, we look for a general
scheme to find the solutions to the (anti-)self-dual equation
\eqref{eq:self-dual_R8_2}.
To this end, we follow the ADHM construction of instantons in four dimensions and generalize it to eight
dimensions.

\subsection{(Anti-)self-dual basis in eight dimensions}
The first step toward the ADHM construction in eight dimensions is to
find an appropriate basis which guarantees the (anti-)self-duality
nature of the gauge field strength $F_{\mu \nu}$.
The corresponding basis in four dimensions is the quaternions
$\sigma_{\mu} = (- i \vec{\sigma}, \mathbf{1}_2)$ 
$(\mu = 1, \ldots, 4)$ where $\vec{\sigma}$ are the Pauli
matrices. Using this basis, quantities $\eta^{(+)}_{\mu\nu} =
\sigma^{\dagger}_{\mu} \sigma_{\nu} - \sigma^{\dagger}_{\nu} \sigma_{\mu}$,
$\eta^{(-)}_{\mu \nu} =
\sigma_{\mu} \sigma^{\dagger}_{\nu} - \sigma_{\nu}
\sigma^{\dagger}_{\mu}$ that satisfy the (anti-)self-dual relations
in four dimensions $\eta^{(\pm)}_{\mu \nu} = \pm \frac{1}{2!}
\varepsilon_{\mu \nu \rho \sigma} \eta^{(\pm)}_{\rho \sigma}$ are defined.
These $\eta^{(\pm)}_{\mu \nu}$ are just the 't Hooft symbol.

By the analogy of the quaternions in four dimensions, we define the
following basis in eight dimensions:
\begin{align}
e_{\mu} = \delta_{\mu 8} \mathbf{1}_8 + \delta_{\mu i} \Gamma_i^{(-)}, \qquad
e_{\mu}^{\dagger} = \delta_{\mu 8} \mathbf{1}_8 + \delta_{\mu
 i}\Gamma_i^{(+)}, 
\qquad 
(\mu = 1, \ldots, 8, \ i= 1, \ldots, 7),
\label{eq:basis}
\end{align}
where $\Gamma_i^{(\pm)}$ are $8 \times 8$ matrices that satisfy the relations $\{ \Gamma_i^{(\pm)}, \Gamma_j^{(\pm)} \} = -2\delta_{ij} \mathbf{1}_8$.
The matrices $\Gamma_i^{(\pm)}$ are defined by $\Gamma_i^{(\pm)} = \frac{1}{2}(1\pm\omega)\Gamma_i$.
We choose the matrices $\Gamma^{(\pm)}_i$ such that they satisfy the relation $\Gamma_i^{(+)}=-\Gamma_i^{(-)}$.
Here $\Gamma_i$ are given by the matrix representation of the seven-dimensional complex (real)
Clifford algebra $\Gamma_i\in C\ell_7\left(\mathbb{C}(\mathbb{R})
\right)$ and $\omega$ is a chirality matrix defined in Appendix \ref{sec:Clifford_ASD-tensor}.
Using this basis, we construct the eight-dimensional counterpart of
the 't Hooft symbol. This is defined by
\begin{align}
\Sigma_{\mu\nu}^{(+)} = e_{\mu}^{\dagger}e_{\nu} -
 e_{\nu}^{\dagger}e_{\mu},
\qquad
\Sigma_{\mu\nu}^{(-)} = e_{\mu}e_{\nu}^{\dagger} -
 e_{\nu}e_{\mu}^{\dagger}.
\label{eq:ASD_tensor_R8_1}
\end{align}
We can confirm that $\Sigma^{(\pm)}_{\mu \nu}$ given
above indeed satisfy the (anti-)self-dual relations
in eight dimensions:
\begin{align}
\Sigma^{(\pm)}_{[\mu\nu}\Sigma^{(\pm)}_{\rho\sigma]} =
 \pm\frac{1}{4!}\varepsilon_{\mu\nu\rho\sigma\alpha\beta\gamma\delta}
\Sigma^{(\pm)}_{\alpha\beta}\Sigma^{(\pm)}_{\gamma\delta},
\label{eq:self-dual_tensor_R8_1}
\end{align}
where the upper script sign of $\Sigma^{(\pm)}_{\mu\nu}$ correspond the sign in the right-hand side of \eqref{eq:self-dual_tensor_R8_1}.
We also observe that the basis $e_{\mu}$ satisfies the following useful relations:
\begin{subequations}
 \begin{align}
e_{\mu}e_{\nu}^{\dagger}+e_{\nu}e_{\mu}^{\dagger} &=
  e^{\dagger}_{\mu}e_{\nu}+e^{\dagger}_{\nu}e_{\mu}
= 2\delta_{\mu\nu} \mathbf{1}_8,
\label{lemm:R8_ASD_algebra_prop_1} \\
e_{\mu}e_{\nu}+e_{\nu}e_{\mu} &=
  2\delta_{\mu8}e_{\nu}+2\delta_{\nu8}e_{\mu}-2\delta_{\mu\nu} \mathbf{1}_8,
\label{lemm:R8_ASD_algebra_prop_2} \\
e_{\mu}^{\dagger}e_{\nu}^{\dagger}+e_{\nu}^{\dagger}e_{\mu}^{\dagger} &=
  2
  \delta_{\mu8}e_{\nu}^{\dagger}+2\delta_{\nu8}e_{\mu}^{\dagger}-2\delta_{\mu\nu}
  \mathbf{1}_8.
\label{lemm:R8_ASD_algebra_prop_3}
 \end{align}
\end{subequations}
Furthermore the basis $e_{\mu}$ is normalized as
$\text{Tr}\left[e_{\mu}e_{\nu}^{\dagger}\right]=8\delta_{\mu\nu}$.
For later convenience we calculate the following quantities:
\begin{align}
& \text{Tr}\Sigma^{(\pm)}_{12}\Sigma^{(\pm)}_{34}\Sigma^{(\pm)}_{56}\Sigma^{(\pm)}_{78}
= \pm 16\text{Tr}\mathbf{1}_8 = \pm128,
\notag \\
& \Sigma^{(\pm)}_{\mu\nu}\Sigma^{(\pm)}_{\rho\sigma}\Sigma^{(\pm)}_{\alpha\beta}
\Sigma^{(\pm)}_{\gamma\delta}  =
  \varepsilon_{\mu\nu\rho\sigma\alpha\beta\gamma\delta}\Sigma^{(\pm)}_{12}
\Sigma^{(\pm)}_{34}\Sigma^{(\pm)}_{56}\Sigma^{(\pm)}_{78}
  = \pm 16 \varepsilon_{\mu\nu\rho\sigma\alpha\beta\gamma\delta}
  \mathbf{1}_8.
\label{eq:R8-ASD_tensor_prop_tr_2_1}
\end{align}

A comment is in order. One may consider that a natural candidate of the
eight-dimensional counterpart of the quaternions is octonions.
Indeed, an ADHM construction of (anti-)self-dual instantons with the octonion
basis has been proposed and studied \cite{Buniy:2002jw, Semikhatov:1985yv}.
However, due to the well-known nature of octonions, the gauge field loses the
associativity which would causes potential difficulties in field
theories. We stress that the basis in \eqref{eq:basis} is defined by the
complex (real) Clifford algebra $C\ell_7 (\mathbb{C} (\mathbb{R}))$ which
has the matrix representations and keeps the associativity.

\subsection{Solutions for gauge field}
Now we have obtained the appropriate basis $e_{\mu}$ which
supplants the quaternions in four dimensions. The next step is to find
explicit solutions for the gauge field $A_{\mu}$.
In the following, we choose the minus sign in \eqref{eq:self-dual_R8_2}
and concentrate on the anti-self-dual equation.
In order to find the anti-self-dual solution,
we first introduce the eight-dimensional Weyl operator\footnote{
Here the symbol $\otimes$ means the Kronecker product,
{\it i.e.} the tensor product of matrices.}
\begin{equation}
\Delta = C(x\otimes\mathbf{1}_k)+D,
\label{eq:R8_ADHM_Weyl_operator}
\end{equation}
where $C$ and $D$ are $(8+8k)\times 8k$ matrices
, $x = x^{\mu} e_{\mu}$ and $x^{\mu}$ are the Cartesian
coordinates of the eight-dimensional Euclid space.
If we consider self-dual solutions, we choose the basis
$e_{\mu}^{\dagger}$ instead of $e_{\mu}$.
The components of the matrices $C$ and $D$ are called the ADHM data.
Note that we can decompose the indices of an $8(1+k)\times 8k$ matrix
into the instanton index that runs from $1$ to $k$ and the color
indices that run from $1$ to $8$.
As we will show, the integer $k$ corresponds to the instanton number
defined by the fourth Chern number $k = \mathcal{N} \int
\mathrm{Tr} [F \wedge F \wedge F \wedge F]$.
Now we introduce an $(8+8k)\times8$ matrix $V(x)$
which satisfies the Weyl equation:
\begin{equation}
\Delta^{\dagger}V(x)=0.
\label{eq:8-dim_Weyl-eq}
\end{equation}
The matrix $V(x)$,
which is called the zero-mode,
is normalized as
\begin{equation}
V^{\dagger}V=\mathbf{1}_8,
\label{eq:8-dim_zero_mode_normalization}
\end{equation}
The completeness condition of $V(x)$ implies
the following relation
\footnote{
The completeness relation \eqref{eq:R8_Weyl_operater_completeness_relation} is derived by assuming an existence of $(\Delta^{\dagger}\Delta)^{-1}$.
}:
\begin{align}
\mathbf{1}_{8+8k}-VV^{\dagger} =
\Delta(\Delta^{\dagger}\Delta)^{-1}\Delta^{\dagger}.
\label{eq:R8_Weyl_operater_completeness_relation}
\end{align}
Following the ADHM construction of instantons in four-dimensions \cite{Atiyah:1978ri}, we
employ the ansatz that the gauge field $A_{\mu} (x)$ is given by the
pure gauge form:
\begin{equation}
A_{\mu}(x) = V^{\dagger}(x)\partial_{\mu}V(x).
\label{eq:R8_ADHM_gauge_zero_mode}
\end{equation}

Next we calculate the field strength $F_{\mu\nu}$ from the ansatz \eqref{eq:R8_ADHM_gauge_zero_mode}.
Using the Weyl equation \eqref{eq:8-dim_Weyl-eq} and the completeness relation \eqref{eq:R8_Weyl_operater_completeness_relation},
the result is
\begin{equation}
F_{\mu\nu} =
 V^{\dagger}C(e_{\mu}\otimes\mathbf{1}_k)(\Delta^{\dagger}\Delta)^{-1}(e_{\nu}^{\dagger}\otimes\mathbf{1}_k)C^{\dagger}V
 - (\mu \leftrightarrow \nu).
\label{eq:R8_ADHM_F_Weyl_zeromode_representation_2}
\end{equation}
We are now looking for conditions that
the field strength
\eqref{eq:R8_ADHM_F_Weyl_zeromode_representation_2} satisfies the anti-self-dual
equation \eqref{eq:self-dual_R8_2}.
One realizes that the basis $e_{\mu}$ should appear in the combination of
$\Sigma^{(-)}_{\mu \nu}$ defined in \eqref{eq:ASD_tensor_R8_1}.
We then demand that the factor $(\Delta^{\dagger} \Delta)^{-1}$ in
\eqref{eq:R8_ADHM_F_Weyl_zeromode_representation_2} commutes with the
basis $e_{\mu}(\otimes\mathbf{1}_k)$:
\begin{align}
e_{\mu}\otimes\mathbf{1}_k(\Delta^{\dagger}\Delta)^{-1} =
(\Delta^{\dagger}\Delta)^{-1}e_{\mu}\otimes\mathbf{1}_k.
\label{eq:ADHM1}
\end{align}
Then the product of the field strengths is calculated to be
\begin{align}
F_{\mu\nu}F_{\rho\sigma} =
\left( V^{\dagger}C(\Delta^{\dagger}\Delta)^{-1}
 \left( \Sigma^{(-)}_{\mu\nu}\otimes\mathbf{1}_k \right) C^{\dagger}V \right) \left( V^{\dagger}C
 \left( \Sigma^{(-)}_{\rho\sigma}\otimes\mathbf{1}_k \right) (\Delta^{\dagger}\Delta)^{-1} C^{\dagger}V
 \right).
\label{eq:FF}
\end{align}
In order that the field strength
$F_{[\mu \nu} F_{\rho\sigma]}$
satisfies the anti-self-dual
equation, 
$\Sigma^{(-)}_{\mu \nu}\otimes\mathbf{1}_k$ should commute with 
$(C^{\dagger} V V^{\dagger} C)$ in \eqref{eq:FF}.
Therefore we demand the following condition:
\begin{align}
e_{\mu}\otimes\mathbf{1}_k\left( C^{\dagger}VV^{\dagger}C \right) = \left(
 C^{\dagger}VV^{\dagger}C \right) e_{\mu}\otimes\mathbf{1}_k.
\label{eq:ADHM2}
\end{align}
Indeed, using the condition \eqref{eq:ADHM2},
the product of the field strengths becomes
\begin{align}
F_{[\mu \nu} F_{\rho \sigma]} = V^{\dagger} C (\Delta^{\dagger}
 \Delta)^{-1}
\left(
\Sigma^{(-)}_{[\mu \nu} \Sigma^{(-)}_{\rho \sigma]}\otimes\mathbf{1}_k
\right)
C^{\dagger} V V^{\dagger} C (\Delta^{\dagger} \Delta)^{-1} C^{\dagger} V.
\label{eq:FF1}
\end{align}
Since $\Sigma^{(-)}_{[\mu\nu} \Sigma^{(-)}_{\rho\sigma]}$ satisfies
the anti-self-dual relation \eqref{eq:self-dual_tensor_R8_1},
we find that this is also true for $F_{[\mu \nu} F_{\rho \sigma]}$.
Therefore the expression \eqref{eq:R8_ADHM_gauge_zero_mode} with the constraints \eqref{eq:ADHM1} and
\eqref{eq:ADHM2} gives the solution to the anti-self-dual equation
\eqref{eq:self-dual_R8_2} in eight dimensions.

It is desirable to find conditions on the ADHM data $C$ and $D$
corresponding to \eqref{eq:ADHM1} and \eqref{eq:ADHM2}.
The equation \eqref{eq:ADHM1} is equivalent to the following constraint
on the matrix $\Delta$:
\begin{align}
\Delta^{\dagger}\Delta  = \mathbf{1}_8 \otimes E^{(1)}_k,
\label{eq:ADHM_c1}
\end{align}
where $E^{(1)}_k$ is an invertible $k \times k$ matrix.
We call \eqref{eq:ADHM_c1} the first ADHM constraint.
The condition \eqref{eq:ADHM_c1}
is a natural generalization of the ADHM constraint in four dimensions.
See Appendix \ref{sec:4-dim_ADHM_U(2)_gauge} for the four-dimensional counterpart of the constraint.

On the other hand, the relation \eqref{eq:R8_Weyl_operater_completeness_relation} allows us to
rewrite the condition \eqref{eq:ADHM2} as
\begin{align}
e_{\mu}\otimes\mathbf{1}_k\left( C^{\dagger}C \right) = \left( C^{\dagger}C \right) e_{\mu}\otimes\mathbf{1}_k,~~~
e_{\mu}\otimes\mathbf{1}_k\left( C^{\dagger}\Delta(\Delta^{\dagger}\Delta)^{-1}\Delta^{\dagger} C \right) = \left( C^{\dagger}\Delta(\Delta^{\dagger}\Delta)^{-1}\Delta^{\dagger} C \right) e_{\mu}\otimes\mathbf{1}_k.
\label{eq:ADHM2-2}
\end{align}
The first condition
is automatically satisfied when the condition
\eqref{eq:ADHM1} holds.
The second one in \eqref{eq:ADHM2-2} is
essentially the new condition for eight-dimensional anti-self-dual
instantons.
This is equivalent to the constraint
\begin{equation}
C^{\dagger}\Delta(\Delta^{\dagger}\Delta)^{-1}\Delta^{\dagger} C =
 \mathbf{1}_8\otimes E^{(2)}_k,
\label{eq:8-dim_ADHM_additional_condition}
\end{equation}
where $E^{(2)}_k$ is an invertible $k \times k$ matrix.
We call \eqref{eq:8-dim_ADHM_additional_condition} the second ADHM constraint.

It is easy to find that the Weyl equation \eqref{eq:8-dim_Weyl-eq}, 
the normalization condition \eqref{eq:8-dim_zero_mode_normalization}, 
the first and the second ADHM constraints 
\eqref{eq:ADHM_c1}, \eqref{eq:8-dim_ADHM_additional_condition} are
invariant under the following transformations:
\begin{align}
C\mapsto C'=\mathcal{U}C\mathcal{R},~~D\mapsto
 D'=\mathcal{U}D\mathcal{R},~~V\mapsto V'=\mathcal{U}V,
\label{eq:ADHM-data_freedom_pre_canonical}
\end{align}
where
$\mathcal{U}\in\text{U}(8+8k)$ and $\mathcal{R}=\mathbf{1}_8\otimes\mathcal{R}_k\in\mathbf{1}_8\otimes\text{GL}(k;\mathbb{C})$
for $\Gamma_i\in C\ell_7(\mathbb{C})$ 
\footnote{
When $\Gamma_i$ take value in the real Clifford algebra $Cl_7
(\mathbb{R})$ instead of the complex one and the ADHM data $C$ are $D$ are real valued,
then the transformation groups are 
$\mathcal{U}\in\text{O}(8+8k)$ and $\mathcal{R}\in 
\mathbf{1}_8\otimes \text{GL}(k;\mathbb{R})$.
}.
Using this $\text{U}(8+8k) \times \text{GL}(k,\mathbb{C})$
transformation, 
we can fix the ADHM data to the so-called canonical form.
This is given by
\begin{align}
C =
\begin{pmatrix}
0_{[8]\times[8k]} \\
\mathbf{1}_{8k}
\end{pmatrix}_{[8+8k]\times[8k]}
, \qquad
D =
\begin{pmatrix}
S_{[8]\times[8k]} \\
T_{[8k]\times[8k]}
\end{pmatrix}_{[8+8k]\times[8k]}
=
\begin{pmatrix}
S_{[8]\times[8k]} \\
e_{\mu~[8]}\otimes T^{\mu}_{[k]}
\end{pmatrix}
.
\label{eq:canonical}
\end{align}
Here the matrix subscript $[a]\times[b]$ means the matrix size.
The symbol $S_{[8]\times[8k]}$ stands for
$\begin{pmatrix}S_{1~[8]\times[k]} &
S_{2~[8]\times[k]} & \dots & S_{8~[8]\times[k]}
\end{pmatrix}$.
We note that all the ADHM data are included in the $(8+8k)\times8k$
matrix $D$ in the canonical form.
We find that there are residual symmetries which leave the canonical
form \eqref{eq:canonical} invariant.
The transformations are given by
\begin{equation}
S_{\mu}\mapsto S'_{\mu}= QS_{\mu}R,~~~T^{\mu}\mapsto T'^{\mu}= R^{\dagger}T^{\mu}R,
\end{equation}
where $Q\in\text{SU}(8)$ and $R\in\text{U}(k)$ for $\Gamma_i\in C\ell_7(\mathbb{C})$
\footnote{When $\Gamma_i\in C\ell_7(\mathbb{R})$ and the matrices $S$
are $T$ are real valued, the transformation groups are
$Q\in\text{SO}(8)$ and $R\in\text{O}(k)$.}.

Now we have established the ADHM construction of (anti-)self-dual
instantons in eight dimensions.
Plugging the canonical form of $C$ and $D$ in \eqref{eq:canonical} into the
first and the second ADHM constraints \eqref{eq:ADHM_c1},
\eqref{eq:8-dim_ADHM_additional_condition}, we obtain the algebraic constraints
on the matrices $T$ and $S$. The explicit form of the constraints (that
are called the ADHM equations) are found in Appendix \ref{sec:R8_U8-ADHM-eq}.
Solutions $S$ and $T$ to these constraints lead to the profile functions of
the gauge field $A_{\mu}$ corresponding to the anti-self-dual instantons.
We will show the explicit solutions for $S$ and $T$ and its associated
gauge field $A_{\mu}$ in Section \ref{sec:R8_ADHM_data}.
However, before going to the solutions, we discuss the gauge groups of
the theory and the homotopy group which classify the solutions.

\subsection{Gauge and homotopy groups} \label{subsec:gauge_homotopy_group}
In this subsection, we discuss the gauge group of the theory and the
homotopy class of the solutions.

The gauge transformation of the solution $A_{\mu}$ is induced by the
transformation of the zero-mode $V(x)$ which preserves 
the normalization condition \eqref{eq:8-dim_zero_mode_normalization}.
Indeed, using the ansatz \eqref{eq:R8_ADHM_gauge_zero_mode}, the
transformation of the zero-mode $V\mapsto Vg(x)$ induces the following
gauge transformation:
\begin{equation}
A_{\mu}\mapsto g^{-1}(x)A_{\mu}g(x)+g^{-1}(x)\partial_{\mu}g(x).
\end{equation}
We note that the transformation $V \mapsto Vg(x) $ is independent of the
one in \eqref{eq:ADHM-data_freedom_pre_canonical}.
The gauge group is determined as follows.

As we have mentioned, the group structure of the transformation matrix
$g(x)$ is determined by the Clifford algebra which has been used to
construct the basis $e_{\mu}$.
For example, when $e_{\mu}$ takes complex values,
then $\Gamma_i$ is the element of the complex Clifford algebra
$C\ell_7(\mathbb{C})$.
In this case, the Weyl operator $\Delta$ takes complex values and 
the solutions to the Weyl equation $\Delta^{\dagger}V=0$ (that is the
zero-mode $V$) is a complex $(8+8k)\times 8$ matrix.
Therefore the gauge group associated with the transformation $V \mapsto Vg(x) $
is the unitary group $G=\text{U}(8)$.
On the other hand, when $e_{\mu}$ and the ADHM data take real values,
then $\Gamma_i$ belongs to the real Clifford algebra $C\ell_7(\mathbb{R})$.
The Weyl operator $\Delta$ takes real values and the zero-mode $V$
is a real $(8+8k)\times 8$ matrix.
In this case, the gauge group associated with the transformation $V \mapsto Vg(x) $
is the orthogonal group $G=\text{O}(8)$.

It is clear that the color size $N$ of the gauge group depends on the matrix size of the basis $e_{\mu}$.
Here the matrix representations of the complex (real) Clifford algebra
are given by the $8\times8$ complex (real) matrices.
Therefore the basis $e_{\mu}$ are $8\times8$ matrices and the color size
is eight, {\it i.e.} $N=8$.
Relations of gauge groups and Clifford algebras are discussed in detail
in Appendix \ref{sec:Clifford_ASD-tensor}.
Note that the ADHM construction does not impose the 
specialty condition on the gauge group
in general, namely, the gauge group $G$ is not 
the special unitary group SU$(N)$ nor the special orthogonal group
SO$(N)$ but they are U$(N)$ or O$(N)$.
We can decompose the group $\text{U}(N)$ (or O$(N)$) into the special
group $\text{SU}(N)$ (or $\text{SO}(N)$) part and $\text{U}(1)$ (or $S^0$)
part: $\text{U}(N)=\text{SU}(N)\ltimes\text{U}(1)$ and $\text{O}(N)=\text{SO}(N)\ltimes S^0$.
Usually, we have to fix the element of $\text{U}(1)$ (or $S^0$) by hand
when we consider SU$(N)$ or SO$(N)$ in the ADHM construction of instantons.

Finally, we give a brief discussion on the homotopy group.
Instantons with gauge group\footnote{Here we focus on the compact Lie
group $G$.} $G$ in eight dimensions are classified by the homotopy group
$\pi_7 (G)$.
We are interested in instantons that are characterized by an integer $k$.
One observes that the gauge group $G$ whose rank is small makes $\pi_7
(G)$ be trivial. For example, the homotopy groups $\pi_7 (G)$ for $G =
\text{SO}(N) \ (N \le 4)$ and $G= \text{SU}(N) \ (N \le 3)$ become trivial.
For larger rank groups, one obtains desired property $\pi_7 (G) =
\mathbb{Z}$ for $G = \text{SU} (N) \ (N \ge 4)$, $G= \text{SO}(N) \ (N \ge 5, N
\not=8)$, $G=\text{Sp}(N) \ (N \ge 2)$.
The homotopy groups relevant to the
eight-dimensional ADHM construction presented in this paper
are $G = \text{U}(8)$, $G = \text{SU}(8)$ and $G=\text{SO}(8)$. For the former two groups, we have
\begin{align}
\pi_7(\text{U}(8))=\pi_7(\text{SU}(8)) = \mathbb{Z},
\end{align}
while for $G = \text{SO}(8)$, we have
\begin{align}
\pi_7 (\text{SO}(8)) = \mathbb{Z} \oplus \mathbb{Z}.
\end{align}
We note that the $\text{SO(}8)$ instanton solutions are embedded in
solutions for the gauge groups $\text{SO}(N) \ (N \ge 8)$.
This is because the property of the homotopy class $\pi_7 (\text{SO}(N)) =
\mathbb{Z} \ (N > 8)$.
The same is true for $\text{SU}(N)$ and $\text{U}(N)$.

\section{ADHM data and multi-instanton solutions} \label{sec:R8_ADHM_data}
In this section, we introduce explicit ADHM data that satisfy the first and the second
ADHM constraints \eqref{eq:ADHM_c1}, \eqref{eq:8-dim_ADHM_additional_condition}.
We will show that the integer $k$ in the construction is
the topological charge of the eight-dimensional instantons.
The topological charge $Q$ for the eight-dimensional instantons is
defined by the fourth Chern number:
\begin{equation}
Q = \mathcal{N}\int_{\mathbb{R}^8}\text{Tr}\left( F\wedge F\wedge F\wedge F \right)
= \mathcal{N} \int_{\mathbb{R}^8} d^8x~\text{Tr}\left(
\left( \frac{1}{2} \right)^4\varepsilon_{\mu\nu\rho\sigma\alpha\beta\gamma\delta}F_{\mu\nu}F_{\rho\sigma}F_{\alpha\beta}F_{\gamma\delta}
 \right),
\label{eq:R8_instanton_charge_origin}
\end{equation}
where $\mathcal{N}$ is the normalization constant which will be determined later.
Using the expression \eqref{eq:FF1} and the ADHM constraints
\eqref{eq:ADHM_c1}, \eqref{eq:8-dim_ADHM_additional_condition}, the charge
density $\mathcal{Q}$ is calculated to be
\begin{equation}
\mathcal{Q} = \pm8!\text{Tr}\left(
V^{\dagger}C(\Delta^{\dagger}\Delta)^{-1}C^{\dagger}V
\right)^4.
\label{eq:R8_ADHM_charge_ADHM_data_represetation_formula_1}
\end{equation}
Here $\pm$ corresponds to the (anti-)self-dual solutions respectively.

In the next subsection, we introduce explicit ADHM data
and calculate the topological charges associated with the solutions.
We first introduce the eight-dimensional ADHM ansatz for the ADHM
data on the analogy of the four-dimensional ones.
Here the ``ansatz'' means that this ADHM data at least satisfy the first ADHM
constraint \eqref{eq:ADHM_c1}.

For a technical reason, it is convenient to introduce the following form of the second
ADHM constraint:
\begin{equation}
C^{\dagger}VV^{\dagger}C = \mathbf{1}_8\otimes E_k^{(3)}, 
\label{eq:R8_pre-additional_ADHM_constraint}
\end{equation}
where $E_k^{(3)}$ is an invertible $k\times k$ matrix.
This is a stronger condition of 
the second ADHM constraint
but more tractable than \eqref{eq:8-dim_ADHM_additional_condition}.
The second ADHM constraint \eqref{eq:8-dim_ADHM_additional_condition} is satisfied when 
\eqref{eq:R8_pre-additional_ADHM_constraint} is satisfied.
In the following, we will examine the second ADHM constraint
\eqref{eq:R8_pre-additional_ADHM_constraint} for given 
ansatz for
ADHM data and determine the multi-instanton profiles.

\subsection{BPST type one-instanton}
We first reproduce the $k=1$ instanton solution in eight dimensions.
This is known as the $\text{SO}(8)$ instanton \cite{Grossman:1984pi}.
In the case of $k=1$,
the ADHM ansatz in the canonical form is taken to be
\begin{align}
&C = \begin{pmatrix}
0\\
\mathbf{1}_8
\end{pmatrix},\hspace{20pt}
D = \begin{pmatrix}
\lambda \mathbf{1}_8 \\
-a^{\mu}e_{\mu}
\end{pmatrix},
\label{eq:R8_ADHM-data_1-instanton}
\end{align}
where $\lambda\in\mathbb{R}$ is the size modulus
and $a^{\mu}\in\mathbb{R}$ is the position modulus of the instanton.
It is easily shown that the ADHM ansatz 
\eqref{eq:R8_ADHM-data_1-instanton} satisfies the first ADHM constraint \eqref{eq:ADHM_c1}.

The solution to the Weyl equation \eqref{eq:8-dim_Weyl-eq} associated
with the ADHM ansatz \eqref{eq:R8_ADHM-data_1-instanton} is found to be
\begin{equation}
V = \frac{1}{\sqrt{\rho}}
\begin{pmatrix}
\tilde{x}^{\dagger} \\ -\lambda\mathbf{1}_8
\end{pmatrix},
\end{equation}
where we have defined $\tilde{x}^{\dagger}=(x^{\mu}-a^{\mu})e^{\dagger}_{\mu}$,
$\|\tilde{x}\|^2=\tilde{x}\tilde{x}^{\dagger}=\tilde{x}^{\dagger}\tilde{x}=
(x^{\mu}-a^{\mu})(x_{\mu}-a_{\mu})$ and $\rho=\lambda^2+\| \tilde{x}
\|^2$.
We next examine the constraint \eqref{eq:R8_pre-additional_ADHM_constraint} for the ADHM ansatz
\eqref{eq:R8_ADHM-data_1-instanton}.
We find that the left-hand side of
\eqref{eq:R8_pre-additional_ADHM_constraint}
associated with the ADHM ansatz \eqref{eq:R8_ADHM-data_1-instanton}
is proportional to the identity $\mathbf{1}_8$:
\begin{align}
C^{\dagger}VV^{\dagger}C = \frac{\lambda^2}{\rho}\mathbf{1}_8.
\end{align}
Therefore, the second ADHM constraint \eqref{eq:8-dim_ADHM_additional_condition} is trivially satisfied.
Then the one-instanton solution to the anti-self-duality equation
in eight dimensions is found to be
\begin{align}
A_{\mu}
= -\frac{1}{2}\frac{x^{\nu}-a^{\nu}}{\lambda^2+\|\tilde{x}\|^2}\Sigma^{(-)}_{\mu\nu}.
\label{eq:BPST-1instanton_ADHM_gauge}
\end{align}
This solution is nothing but the $\text{SO}(8)$ instanton found in
\cite{Grossman:1984pi}. 
This is the eight-dimensional analogue of the
Belavin-Polyakov-Schwarz-Tyupkin (BPST) instanton \cite{Belavin:1975fg}
in four dimensions.
The associated field strength $F_{\mu\nu}$ is evaluated to be
\begin{equation}
F_{\mu\nu} = \partial_{\mu}A_{\nu}-\partial_{\nu}A_{\mu} +
 [A_{\mu},A_{\nu}] =
 \frac{\lambda^2}{(\lambda^2+\|\tilde{x}\|^2)^2}\Sigma^{(-)}_{\mu\nu}.
\label{eq:BPST-1instanton_ADHM_field_strength}
\end{equation}
Then the ADHM construction in eight dimensions have reproduced the known
one-instanton solution.
Next we calculate the topological charge and determine the normalization
constant $\mathcal{N}$.
The field strength for the $\text{SO}(8)$ instanton
\eqref{eq:BPST-1instanton_ADHM_field_strength} is very simple, so we are
able to calculate the charge using
\eqref{eq:R8_instanton_charge_origin}.
The result is
\begin{align}
Q &= \mathcal{N}\frac{1}{2^4}8!\int_{\mathbb{R}^8} d^8x~\left( \frac{\lambda^2}{(\lambda^2+\tilde{x}^2)^2}\right)^4\text{Tr}\left( \Sigma^{(-)}_{12}\Sigma^{(-)}_{34}\Sigma^{(-)}_{56}\Sigma^{(-)}_{78} \right)
= -384\pi^4\mathcal{N}.\label{eq:1-inst_charge}
\end{align}
Therefore the normalization constant $\mathcal{N}$ is
determined to be
\begin{equation}
\mathcal{N}=\frac{1}{4!(2\pi)^4}.
\end{equation}
This normalization is the same one employed in \cite{Grossman:1984pi}.

\subsection{'t Hooft type solutions}
We next study ADHM data for instantons with $k \ge 2$.
A natural candidate for this is 
an eight-dimensional
 generalization of the 't Hooft type one \cite{Atiyah:1978ri}.
The 't Hooft type ADHM ansatz are given by
\begin{align}
T^{\mu} &= \text{diag}_{p=1}^k\left( -a^{\mu}_p \right),
\notag \\
S &=\mathbf{1}_8 \otimes
\begin{pmatrix}
\lambda_1 & \lambda_2 & \dots & \lambda_k
\end{pmatrix},
\label{eq:R8-ADHM_data_tHooft_k-instanton_data0}
\end{align}
where $a^{\mu}_p\in\mathbb{R}$ are position and $\lambda_p\in\mathbb{R}$ are size moduli respectively.
The Weyl operator associated with the 't Hooft type ADHM ansatz is
\begin{equation}
\Delta^{\dagger} =
\begin{pmatrix}
S^{\dagger} & e^{\dagger}_{\mu}\otimes(x^{\mu}\mathbf{1}_k + T^{\mu})
\end{pmatrix}.
\end{equation}
Then we find
\begin{equation}
\Delta^{\dagger}\Delta
= \mathbf{1}_8\otimes
\begin{pmatrix}
 \lambda_1^2+\|\tilde{x}_1\|^2 & \lambda_1\lambda_2 & \dots &
 \lambda_1\lambda_k \\
 \lambda_2\lambda_1 & \lambda_2^2+\|\tilde{x}_2\|^2 & \dots &
 \lambda_2\lambda_k \\
 \vdots & \vdots & \ddots & \vdots \\
 \lambda_k\lambda_1 & \lambda_k\lambda_2 & \dots & \lambda_k^2+\|\tilde{x}_k\|^2
\end{pmatrix}.
\label{eq:tH_ADHM}
\end{equation}
Here $\tilde{x}_p$ is defined as $\tilde{x}_p=(x^{\mu}-a^{\mu}_p)e_{\mu}$.
Therefore the ADHM ansatz
\eqref{eq:R8-ADHM_data_tHooft_k-instanton_data0} satisfies the first ADHM
constraint \eqref{eq:ADHM_c1}.

The solution to the Weyl equation \eqref{eq:8-dim_Weyl-eq} is given by
\begin{equation}
V=\frac{1}{\sqrt{\phi}}
\begin{pmatrix}
-\mathbf{1}_8 \\
\left( e_{\mu}\otimes\text{diag}_{p=1}^k\left( \frac{\tilde{x}_p^{\mu}}{\|\tilde{x}_p\|^2} \right) \right)S^{\dagger}
\end{pmatrix},
\label{R8-ADHM_data_zero_mode_general}
\end{equation}
where $\phi=1+\sum_{p=1}^k\frac{\lambda_p^2}{\|\tilde{x}_p\|^2}$.
We then examine the constraint \eqref{eq:R8_pre-additional_ADHM_constraint}.
Plugging the zero-mode \eqref{R8-ADHM_data_zero_mode_general} into
$C^{\dagger} V V^{\dagger} C$, we have
\begin{align}
C^{\dagger}VV^{\dagger}C
= \left( \delta_{\mu\nu}\mathbf{1}_8 + \Sigma^{(-)}_{\mu\nu}/2 \right)
 \otimes
E^{\mu \nu}_{(\text{'t Hooft})},
\label{eq:tHooft_ADHM}
\end{align}
where
\begin{align}
E^{\mu \nu}_{(\text{'t Hooft})}
= \begin{pmatrix}
\lambda_1^2X_1^{\mu}X_1^{\nu} & \lambda_1\lambda_2X_1^{\mu}X_2^{\mu} &
 \dots & \lambda_1\lambda_kX_1^{\mu}X_k^{\nu} \\
\lambda_2\lambda_1X_2^{\mu}X_1^{\nu} & \lambda_2^2X_2^{\mu}X_2^{\nu}
 &\dots & \lambda_2\lambda_kX_2^{\mu}X_k^{\nu} \\
\vdots & \vdots & \ddots & \vdots \\
\lambda_k\lambda_1X_k^{\mu}X_1^{\nu} &
 \lambda_k\lambda_2X_k^{\mu}X_2^{\nu} & \dots & \lambda_k^2X_k^{\mu}X_k^{\nu}
\end{pmatrix}.
\end{align}
Here we have used the relation
$e_{\mu}e_{\nu}^{\dagger}=\delta_{\mu\nu}\mathbf{1}_8+\Sigma_{\mu\nu}^{(-)}/2$
and defined $X_m^{\mu}=\tilde{x}_m^{\mu}/\|\tilde{x}_m\|^2$.
Since the constraint \eqref{eq:R8_pre-additional_ADHM_constraint}
requires that the right-hand side of
\eqref{eq:tHooft_ADHM} is proportional to $\mathbf{1}_8$, we have the
following conditions on the moduli $\lambda_a$ and $a^{\mu}_m$:
\begin{equation}
\lambda_m\lambda_n(x^{\mu}-a_m^{\mu})(x^{\nu}-a_n^{\nu})\Sigma^{(-)}_{\mu\nu}=0.
\label{eq:8-dim_tHooft_ADHM_data_additional_condition}
\end{equation}
Here the indices $m,n$ run from 1 to $k$ and not summed.
We find that the conditions \eqref{eq:8-dim_tHooft_ADHM_data_additional_condition}
are satisfied in the well-separated limit of each instanton:
\begin{equation}
\|a_m^{\mu}-a_n^{\mu}\|^2
\gg
\lambda_m\lambda_n,
\label{eq;R8_instantons_well-separated_limit}
\end{equation}
for all $m$ and $n$.
In the well-separated limit
\eqref{eq;R8_instantons_well-separated_limit},
we can neglect all the off-diagonal components in
the matrix in \eqref{eq:tH_ADHM}:
\begin{align}
\begin{pmatrix}
\lambda_1^2+\|\tilde{x}_1\|^2 & \dots & \lambda_1\lambda_k \\
\vdots & \ddots & \vdots \\
\lambda_k\lambda_1 & \dots & \lambda_k^2+\|\tilde{x}_k\|^2
\end{pmatrix}
\simeq
\begin{pmatrix}
 \lambda_1^2+\|\tilde{x}_1\|^2 & \dots & 0 \\
 \vdots & \ddots & \vdots \\
 0 & \dots & \lambda_k^2+\|\tilde{x}_k\|^2
\end{pmatrix}.
\label{eq:R8-tHooft_data_well-separated}
\end{align}
Indeed, in this limit we have
\begin{align}
C^{\dagger}\Delta(\Delta^{\dagger}\Delta)^{-1}\Delta^{\dagger} C
\simeq \mathbf{1}_8\otimes \text{diag}_{p=1}^k\left[
 \frac{\|\tilde{x}_p\|^2}{\lambda_p^2+\|\tilde{x}_p\|^2} \right].
\label{eq:8-dim_well-separated_approximate_additional_ADHM_constraint}
\end{align}
Therefore for the 't Hooft type ansatz
\eqref{eq:R8-ADHM_data_tHooft_k-instanton_data0},
the second ADHM constraint is satisfied in the well-separated
limit.
Since the number of instanton density becomes dilute in this limit, this is called
the dilute instanton gas approximation \cite{Christ:1978jy}.

We proceed to evaluate the instanton charge for the 't Hooft type ADHM
data. In the well-separated limit
\eqref{eq;R8_instantons_well-separated_limit}, 
by using \eqref{eq:R8_ADHM_charge_ADHM_data_represetation_formula_1}, 
the charge density for general $k$ instantons is calculated as
\begin{equation}
\mathcal{Q} \simeq \frac{-8!}{\phi^4}\text{Tr}\left\{
  f^{\mu\nu}e^{\dagger}_{\mu}e_{\nu} \right\}^4,
\label{eq:R8_ADHM_charge_density_tHooft_general_1}
\end{equation}
where $f^{\mu \nu}$ is given by
\begin{equation}
f^{\mu\nu}=
\begin{pmatrix} \frac{\lambda_1\tilde{x}_1^{\mu}}{\|\tilde{x}_1\|^2} & \frac{\lambda_2\tilde{x}_2^{\mu}}{\|\tilde{x}_2\|^2} & \dots & \frac{\lambda_k\tilde{x}_k^{\mu}}{\|\tilde{x}_k\|^2} \end{pmatrix}
\begin{pmatrix}
\lambda_1^2+\|\tilde{x}_1\|^2 & \lambda_1\lambda_2 & \dots &
 \lambda_1\lambda_k \\
\lambda_1\lambda_2 & \lambda_2^2+\|\tilde{x}_2\|^2 & \dots &
 \lambda_2\lambda_k \\
\vdots & \vdots & \ddots & \vdots \\
\lambda_1\lambda_k & \lambda_2\lambda_k & \dots & \lambda_k^2+\|\tilde{x}_k\|^2
\end{pmatrix}^{-1}
\begin{pmatrix} \frac{\lambda_1\tilde{x}_1^{\nu}}{\|\tilde{x}_1\|^2} \\ \frac{\lambda_2\tilde{x}_2^{\nu}}{\|\tilde{x}_2\|^2} \\ \vdots \\ \frac{\lambda_k\tilde{x}_k^{\nu}}{\|\tilde{x}_k\|^2} \end{pmatrix}.
\end{equation}
In order to illustrate multi-instanton solutions, we write down the
charge densities for $k=1,2,3$ explicitly.

For $k=1$, the charge density is given by
\begin{align}
\mathcal{Q}^{(k=1)}_{\text{'t Hooft}} = -8!\cdot8\left( \frac{\lambda^2}{\left( \lambda^2+\|\tilde{x}\|^2
 \right)^2} \right)^4.
\label{eq:R8_1-tHooft_instanton_charge_density}
\end{align}
This is nothing but the one calculated in \eqref{eq:1-inst_charge}.
We note that the one-instanton solution in the 't Hooft ADHM data is
singular at the instanton position:
\begin{equation}
A_{\mu}^{\text{singular}} =
 \frac{1}{4}\Sigma^{(+)}_{\mu\nu}\partial_{\nu}
\ln \left( 1+\frac{\lambda^2}{\|\tilde{x}\|^2} \right),
\label{eq:1-inst_singular}
\end{equation}
while the BPST type solution \eqref{eq:BPST-1instanton_ADHM_gauge}
discussed in the previous subsection is non-singular.
These solutions are connected by the following singular gauge transformation:
\begin{equation}
A_{\mu}^{\text{non-singular}}=g_1A_{\mu}^{\text{singular}}g_1^{-1}+g_1\partial_{\mu}g_1^{-1},
\qquad g_1=\frac{\tilde{x}}{\sqrt{\|\tilde{x}\|^2}}.
\end{equation}
%%%%%%%%%%%%%%%%%%%%%%%%%%%%%%%%%%%
\begin{figure}[tbp]
\includegraphics{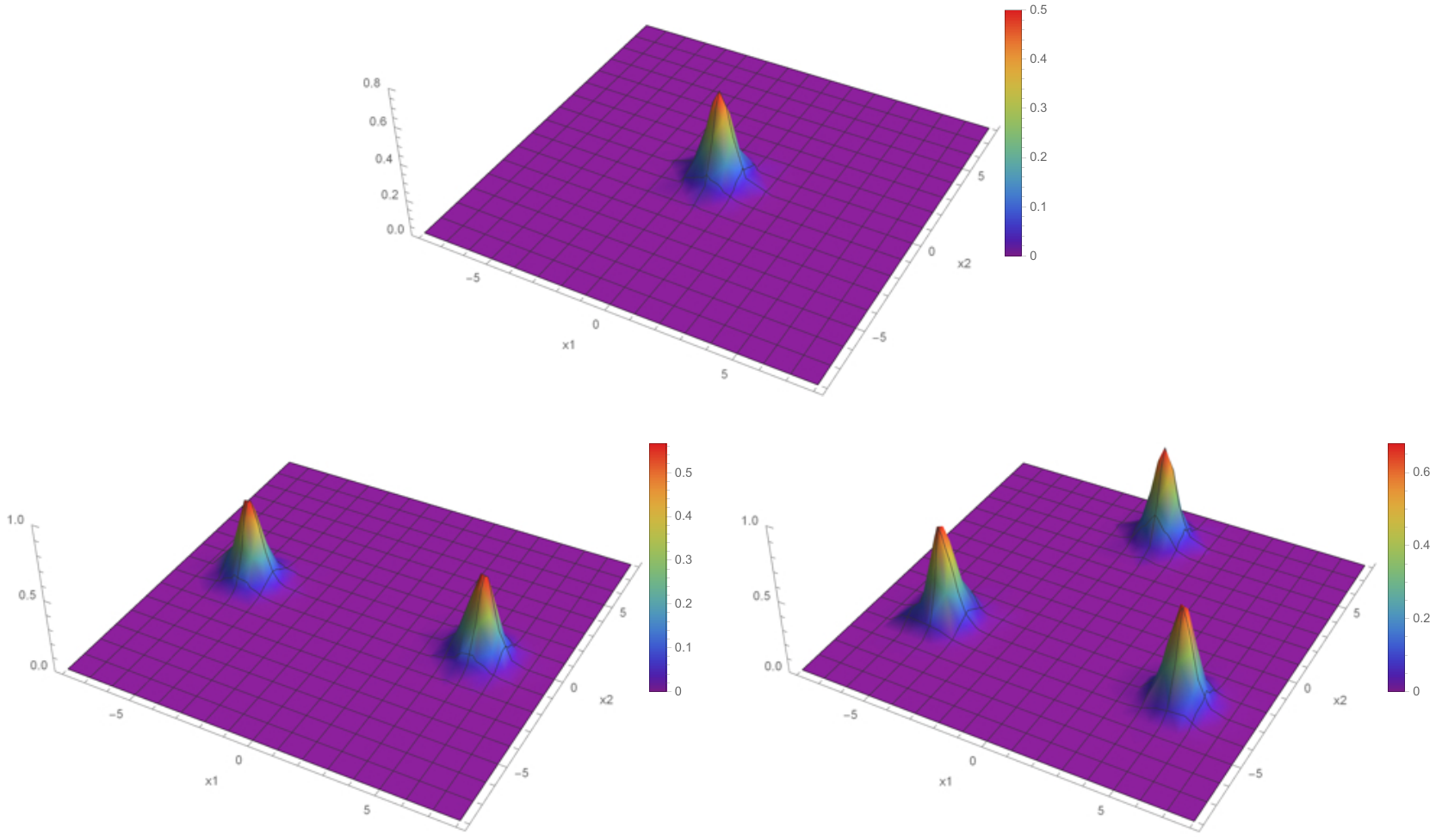}
\caption{
The charge density plots of the 't Hooft type solutions.
The upper figure corresponds to $k=1$, left and right ones 
in the lower figure correspond to $k=2,3$ respectively.
All the plots are projected to a two-dimensional subspace in
the eight-dimensional space.
}
\label{fig:charge_density_tHooft}
\end{figure}
%%%%%%%%%%%%%%%%%%%%%%%%%%%%%%%%%%%
For $k=2$ and $k=3$, the charge densities
in the dilute gas approximation are evaluated as
\begin{align}
\mathcal{Q}^{(k=2)}_{\text{'t Hooft}} \simeq& \  -8!\cdot8\left(
\frac{\lambda_1^2\|\tilde{x}_2\|^4+\lambda_2^2\|\tilde{x}_1\|^4+\lambda_1^2\lambda_2^2\left(
\|\tilde{x}_1\|^2+\|\tilde{x}_2\|^2-2\tilde{x}_1^{\mu}\tilde{x}_2^{\mu}
\right)}{\left(
\lambda_1^2\|\tilde{x}_2\|^2+\lambda_2^2\|\tilde{x}_1\|^2+\|\tilde{x}_1\|^2\|\tilde{x}_2\|^2
\right)^2}
\right)^4
\label{eq:R8_2-instanton_charge_density_singular}
, \\
\mathcal{Q}^{(k=3)}_{\text{'t Hooft}} \simeq& \  -8!\cdot8\Biggl[ \gamma
\biggl( \lambda_1^2\|\tilde{x}_2\|^4\|\tilde{x}_3\|^4 +
 \lambda_2^2\lambda_3^2\|\tilde{x}_1\|^4\left(
 \|\tilde{x}_2\|^2+\|\tilde{x}_3\|^2-2\tilde{x}_2^{\mu}\tilde{x}_3^{\mu}
 \right) \notag \\
&\hspace{70pt} +\lambda_2^2\|\tilde{x}_1\|^4\|\tilde{x}_3\|^4+
 \lambda_1^2\lambda_3^2\|\tilde{x}_2\|^4\left(
 \|\tilde{x}_1\|^2+\|\tilde{x}_3\|^2-2\tilde{x}_1^{\mu}\tilde{x}_3^{\mu}
 \right)
\notag \\
&\hspace{80pt}
 +\lambda_3^2\|\tilde{x}_1\|^4\|\tilde{x}_2\|^4 +
 \lambda_1^2\lambda_2^2\|\tilde{x}_3\|^4\left(
 \|\tilde{x}_1\|^2+\|\tilde{x}_2\|^2-2\tilde{x}_1^{\mu}\tilde{x}_2^{\mu}
 \right)  \biggr)\Biggr]^4.
 \label{eq:R8_3-instanton_charge_density_1}
\end{align}
Here we have defined
\begin{align}
\gamma= \frac{1}{\left(
 \lambda_1^2\|\tilde{x}_2\|^2\|\tilde{x}_3\|^2+\lambda_2^2\|\tilde{x}_1\|^2\|\tilde{x}_3\|^2+\lambda_3^2\|\tilde{x}_1\|^2\|\tilde{x}_2\|^2+\|\tilde{x}_1\|^2\|\tilde{x}_2\|^2\|\tilde{x}_3\|^2
 \right)^2}.
\end{align}

The numerical profiles for the $k=1,2,3$ charge densities are found in
Fig \ref{fig:charge_density_tHooft}.
Here the parameters that satisfy the well-separated limit
\eqref{eq;R8_instantons_well-separated_limit}
are chosen such that
$a^{\mu}=0,~\lambda=2$ for $k=1$,
$a_1^1 = -5 $, $a_2^1 = 5$, $a_1^{\mu}=a_2^{\mu}=0$, $(\mu>1)$, $\lambda_1=\lambda_2=2$,
 for $k=2$ and $a_m^1=10/\sqrt{3}\times \sin(2\pi(m-1)/3)$, $a_m^2=10/\sqrt{3}\times \cos(2\pi(m-1)/3)$, $a_m^{\mu}=0$, $(\mu>2)$,
$\lambda_m=2$, $(m=1,2,3)$ for $k=3$.
For these parameters, the
numerical results of instanton charges are evaluated as
$Q \simeq 2 \times 1.02 \ (k=2)$,
$Q \simeq 3 \times 1.03 \ (k=3)$.
Therefore we find that the dilute instanton gas approximation, which is needed to
solve the second ADHM constraint, works well.

We also observe that the topological charge defined by the fourth Chern number is quantized in the well-separated limit.
Indeed, using the property of the basis $e_{\mu}$, the charge density
formula for general anti-self-dual instantons
\eqref{eq:R8_ADHM_charge_ADHM_data_represetation_formula_1} is rewritten as
\begin{equation}
\mathcal{Q} = -8!\cdot8\text{Tr}_k\left( \left( E^{(1)}_k \right)^{-1} \left( \mathbf{1}_k-E_k^{(2)} \right) \right)^4,
\end{equation}
where ADHM data have been fixed to the canonical form.
For the 't Hooft type ADHM data in the dilute gas approximation, we have
\begin{equation}
\mathcal{Q}_{\text{'t Hooft}} 
\simeq -8!\cdot8\sum_{p=1}^k\left( \frac{\lambda_p^2}{(\lambda_p^2+\|\tilde{x}_p\|^2)^2} \right)^4.
\end{equation}
This is just the summation of the one-instanton charge density \eqref{eq:R8_1-tHooft_instanton_charge_density}
and gives $Q = k$.

A few comments are in order.
First, we find the special solutions to the condition
\eqref{eq:8-dim_tHooft_ADHM_data_additional_condition}.
The condition is exactly solved by $a_m = a_n \ (m \not= n)$ which
implies that all the instantons are localized at the same point.
However, we find that the corresponding solution is equivalent to the
one-instanton \eqref{eq:1-inst_singular}.
On the other hand, another exact solution $\lambda_n = 0$ (for all $n$) make the
solution be trivial
\footnote{
Strictly speaking, the solution becomes singular at the instanton
positions $\tilde{x}^{\mu}_m = 0$ in the limit $\lambda_m \to 0$.
We will discuss the physical meaning of this limit in Section 4.
}
. Namely, it is a vacuum configuration.

Second, there is a principal difference between the four- and the
eight-dimensional (anti-)self-dual equations.
In four dimensions, the 't Hooft type ADHM data provides the {\it exact} solutions to
the (anti-)self-dual equation $F = \pm *_4 F$ \cite{Atiyah:1978ri}.
However, in eight dimensions, this provides only the {\it approximate}
solutions.
The reason is that the (anti-)self-dual equation is linear in $F$ only in four
dimensions.
The (anti-)self-dual equations in dimensions greater than four
contain multiple $F$. For example in $4n \ (n \ge 2)$ dimensions, the
equation is given by
\begin{align}
F^{\wedge n} = \pm *_{4n} F^{\wedge n},
\label{eq:SD_4n}
\end{align}
where $F^{\wedge n}$ is the wedge products of $n$ field strengths $F$.
The equations \eqref{eq:SD_4n} are non-linear in $F$ when $n \ge 2$.
The intrinsic origin of the second ADHM constraint
\eqref{eq:8-dim_ADHM_additional_condition} comes from
this non-linearity of the (anti-)self-dual equations.
Therefore the situation in $4n \ (n \ge 2)$ dimensions is quite
different from the four-dimensional case.

\subsection{Jackiw-Nohl-Rebbi type solutions}
We then study a generalization of the 't Hooft solutions which is
so-called Jackiw-Nohl-Rebbi (JNR) type solutions.

The JNR type ansatz \cite{Corrigan:1978ce} is given by
\begin{equation}
\Delta = \begin{pmatrix}
\mathbf{1}_8\otimes\Lambda \\
\mathbf{1}_8\otimes\mathbf{1}_{k}
\end{pmatrix} \cdot x\otimes\mathbf{1}_k
+ \begin{pmatrix}
-a_0\otimes\Lambda \\
\text{diag}_{p=1}^k(-a_p)
\end{pmatrix}
=
\begin{pmatrix}
\tilde{x}_0\otimes\Lambda	\\
\tilde{X}_{[8k]\times[8k]}
\end{pmatrix}
=e_{\mu}\otimes
\begin{pmatrix}
\tilde{x}_0^{\mu}\Lambda	\\
\text{diag}_{p=1}^k(\tilde{x}_p^{\mu})
\end{pmatrix},	\label{eq:R8_JNR_ADHM_data}
\end{equation}
where
 $\Lambda= \begin{pmatrix}\lambda_1/\lambda_0 & \dots & \lambda_k/\lambda_0\end{pmatrix}$,
$\tilde{x}_i=(x^{\mu}-a_i^{\mu})e_{\mu},~a_i= a_i^{\mu}e_{\mu}$ and
$\tilde{X}=\text{diag}(\tilde{x}_1,\dots,\tilde{x}_k)$.
Here $\lambda_i\in\mathbb{R}$ and $a_i^{\mu}\in\mathbb{R}$
$(i=0,\dots,k)$ are moduli parameters.
We note that the JNR ansatz \eqref{eq:R8_JNR_ADHM_data} is not in
the canonical form and contain more moduli parameters than the 't Hooft
one. The latter is obtained from the former by the limit 
$a_0\to\infty,~\lambda_0\to\infty$ with fixed $a_0/\lambda_0=1$.

We can confirm that the JNR ansatz satisfies the first ADHM constraint \eqref{eq:ADHM_c1}:
\begin{align}
\Delta^{\dagger}\Delta =\mathbf{1}_8\otimes \left( \|\tilde{x}_0\|^2~{}^{t}\Lambda\Lambda+\text{diag}_{p=1}^k(\|\tilde{x}_p\|^2)\right)
=\mathbf{1}_8\otimes E_k^{(\text{JNR})},	\label{eq:R8_JNR_data_standard_ADHM}
\end{align}
where the symbol ${}^tM$ means the transposed matrix of $M$, so
${}^t\Lambda$ is $k$-column vector and ${}^t\Lambda\Lambda$ is $k\times
k$ matrix.
The solution to the Weyl equation \eqref{eq:8-dim_Weyl-eq} is given by
\begin{equation}
V = \frac{1}{\sqrt{\phi}}
\begin{pmatrix}
-\mathbf{1}_8	\\
\text{diag}_{p=1}^k\left( \frac{\tilde{x}_p}{\|\tilde{x}_p\|^2} \right)~\cdot~\tilde{x}_{0}^{\dagger}\otimes{}^t\Lambda
\end{pmatrix},
\end{equation}
where $\phi=1+\frac{\|\tilde{x}_0\|^2}{\lambda_0^2}~\sum_{p=1}^k\left(
\frac{\lambda_p^2}{\|\tilde{x}_p\|^2} \right)$.

Now we examine the second ADHM constraint.
The left-hand side of \eqref{eq:R8_pre-additional_ADHM_constraint} is
evaluated to be:
\begin{eqnarray}
C^{\dagger}VV^{\dagger}C &=&  \frac{1}{\phi\lambda_0^2}\left(
e_{\mu}e_{\nu}^{\dagger}e_{\rho}e_{\sigma}^{\dagger} \otimes
E^{\mu\nu\rho\sigma}_{(\text{JNR})}
\right),
\nonumber \\
E^{\mu\nu\rho\sigma}_{(\text{JNR})}
&=&  
\begin{pmatrix}
\lambda_1^2Y^{\mu\nu}_1Y^{\sigma\rho}_1	& \lambda_1\lambda_2Y^{\mu\nu}_1Y^{\sigma\rho}_2	& \dots & \lambda_1\lambda_kY^{\mu\nu}_1Y^{\sigma\rho}_k	\\
\lambda_2\lambda_1Y^{\mu\nu}_2Y^{\sigma\rho}_1	& \lambda_2^2	Y^{\mu\nu}_2Y^{\sigma\rho}_2&\dots	& \lambda_2\lambda_kY^{\mu\nu}_2Y^{\sigma\rho}_k	\\
\vdots	& \vdots	& \ddots	& \vdots	\\
\lambda_k\lambda_1Y^{\mu\nu}_kY^{\sigma\rho}_1	& \lambda_k\lambda_2Y^{\mu\nu}_kY^{\sigma\rho}_2	& \dots	& \lambda_k^2Y^{\mu\nu}_kY^{\sigma\rho}_k
\end{pmatrix}
,	\label{eq:8-dim_JNR_data_additional_codition}
\end{eqnarray}
where $Y^{\mu\nu}_m= \tilde{x}^{\mu}_m\tilde{x}^{\nu}_{0}\Big/\|\tilde{x}_m\|^2 -\delta^{\mu8}\delta^{\nu8}$ and $m=1,\dots,k$ is not summed.
In each component in the matrix in
\eqref{eq:8-dim_JNR_data_additional_codition}, we have
\begin{equation}
Y^{\mu\nu}_mY^{\sigma\rho}_ne_{\mu}e_{\nu}^{\dagger}e_{\rho}e_{\sigma}^{\dagger}=
\frac{\|\tilde{x}_0\|^2}{\|\tilde{x}_m\|^2\|\tilde{x}_n\|^2}\tilde{x}_m\tilde{x}^{\dagger}_n - \frac{1}{\|\tilde{x}_n\|^2}\tilde{x}_0\tilde{x}^{\dagger}_n - \frac{1}{\|\tilde{x}_m\|^2}\tilde{x}_m\tilde{x}^{\dagger}_0 + \mathbf{1}_8.
\label{eq:R8_JNR_additional_constraint_I}
\end{equation}

For $k=1$, since we have the relation
$\tilde{x}_a\tilde{x}_b^{\dagger}+\tilde{x}_b\tilde{x}_a^{\dagger}=2\tilde{x}_a^{\mu}\tilde{x}_b^{\mu}\mathbf{1}_8$,
the right-hand side of  
\eqref{eq:R8_JNR_additional_constraint_I} is proportional to
$\mathbf{1}_8$ and the second ADHM constraint is satisfied.
The charge density of the $k=1$ JNR solution is given by
\begin{equation}
\mathcal{Q}^{(k=1)}_{\text{JNR}} = -8!\cdot8\left( \frac{ \bar{\lambda}_1^2\left( \|\tilde{x}_1\|^2+\|\tilde{x}_0\|^2-2\tilde{x}_0^{\mu}\tilde{x}_1^{\mu} \right)}{\left( \|\tilde{x}_0\|^2\bar{\lambda}_1^2+\|\tilde{x}_1\|^2 \right)^2} \right)^4, \label{eq:R8_1-JNR_charge_density}
\end{equation}
where $\bar{\lambda}_m=\lambda_m/\lambda_0$.
The moduli parameters are $\lambda_1/\lambda_0=\lambda$ and
$a_1^{\mu}-a_0^{\mu}=a^{\mu}$, so the $k=1$ JNR solution has total nine
parameters.
Therefore the $k=1$ JNR data is essentially equal to the $k=1$ 't Hooft
data, and we find that the numerical results of the $k=1$ instanton charge \eqref{eq:R8_1-JNR_charge_density} is $Q=1$.

For $k\geq2$ case, it is not straightforward to solve the constraint
\eqref{eq:R8_pre-additional_ADHM_constraint} in a general fashion.
However, a solution is found in the well-separated limit
\eqref{eq;R8_instantons_well-separated_limit}. 
In this limit, we can neglect all the off-diagonal components in $E^{(\text{JNR})}_k$:
\begin{equation}
E_k^{(\text{JNR})} \simeq \text{diag}_{p=1}^k\left( \|\tilde{x}_0\|^2\bar{\lambda}_p^2+\|\tilde{x}_p\|^2 \right).
\end{equation}
Then, the second ADHM constraint is satisfied:
\begin{equation}
C^{\dagger}\Delta(\Delta^{\dagger}\Delta)^{-1}\Delta^{\dagger}C \simeq
\mathbf{1}_8\otimes \text{diag}_{p=1}^k\left( \frac{\|\tilde{x}_0\|^2\bar{\lambda}_p^4+2\bar{\lambda}_p^2\tilde{x}_0^{\mu}\tilde{x}_p^{\mu}+\|\tilde{x}_p\|^2}{\|\tilde{x}_0\|^2\bar{\lambda}_p^2+\|\tilde{x}_p\|^2} \right).
\end{equation}

We also observe that the instanton charge is quantized in this limit by
using the same formula of the 't Hooft ones.
We note that the JNR data is not in the canonical form.
In this case, the charge density formula \eqref{eq:R8_ADHM_charge_ADHM_data_represetation_formula_1} is rewritten as
\begin{equation}
\mathcal{Q} = -8!\cdot8\text{Tr}_k\left( \left( E_k^{(1)}\right)^{-1}(C^{(2)}-E_k^{(2)}) \right)^4,
\end{equation}
where $C^{(2)}$ is defined by $C^{\dagger}C=\mathbf{1}_8\otimes C^{(2)}$.
In the limit \eqref{eq;R8_instantons_well-separated_limit}, we have 
$C^{\dagger}C\simeq \mathbf{1}_8\otimes (\text{diag}_{p=1}^k\bar{\lambda}_p^2+\mathbf{1}_k)$.
Therefore we obtain
\begin{equation}
\mathcal{Q}_{\text{JNR}} \simeq
-8!\cdot8\sum_{p=1}^k \left( \frac{ \bar{\lambda}_p^2\left(
 \|\tilde{x}_p\|^2+\|\tilde{x}_0\|^2-2\tilde{x}_0^{\mu}\tilde{x}_p^{\mu}
\right)}{\left(
\|\tilde{x}_0\|^2\bar{\lambda}_p^2+\|\tilde{x}_p\|^2
\right)^2}
\right)^4. 
\label{eq:JNR_charge}
\end{equation}
This is just the summation of the JNR type one-instanton charge density
and the charge associated with \eqref{eq:JNR_charge} is $Q = k$.

We note that these three type ADHM data (BPST type, 't Hooft type and
JNR type) take real values.
Therefore we can choose the gauge group $G$ by using the Clifford algebras: $C\ell_7(\mathbb{C})$ or $C\ell_7(\mathbb{R})$.
If we choose the complex Clifford algebra $C\ell_7(\mathbb{C})$ then the gauge group is the unitary
group $G=\text{U}(8)$. On the other hand, we choose the real Clifford
algebra $C\ell_7(\mathbb{R})$ 
for the orthogonal group $G=\text{O}(8)$.
We find explicit form of the complex (real) basis in Appendix \ref{sec:Clifford_ASD-tensor}.

\section{Higher derivative field theories in eight dimensions}
In this section we discuss eight-dimensional gauge field theories where the
(anti-)self-dual instantons are relevant.
Since the (anti-)self-dual equations in dimensions greater than four contain multi-field
strengths, the theories inevitably contain higher derivative terms.
In the following, we consider a gauge field $A_{\mu}$ and a non-Abelian gauge
group $G$ whose Lie algebra is $\mathcal{G}$ in eight-dimensional Euclid space.
The generators of the gauge group
$T^a \ (a = 1, \ldots, \text{dim} \ \mathcal{G})$ are normalized as
 $\mathrm{Tr} T^a T^b = \kappa \delta^{ab}$ where $\kappa$ is a constant.
We also introduce the gauge coupling constant $g$ whose mass
 dimension is $-2$ in eight dimensions.
The constant $\alpha'$ is the string Regge slope parameter.

\paragraph{Quartic Yang-Mills model}
The first example is the so called quartic Yang-Mills model whose Lagrangian is
given by the 4th products of the gauge field strengths $F$ and no
Yang-Mills kinetic term.
The action of the quartic Yang-Mills model is given by
\begin{align}
S = \frac{\alpha}{\kappa g^2} \int  \mathrm{Tr}
\left[
\frac{1}{2}
*_8 (F \wedge F) \wedge (F \wedge F)
\right],
\label{eq:quartic}
\end{align}
where $\alpha$ is a constant whose mass dimension is $[\alpha] = -4$.
The action \eqref{eq:quartic} is classically conformal and the Derrick's
theorem implies that the theory admits stable static solitons.
It is straightforward to show that the Bogomol'nyi completion of the action is
\begin{align}
S = \frac{\alpha}{\kappa g^2}
\int \! \mathrm{Tr}
\left[
(F \wedge F \pm *_8 (F \wedge F))^2 \mp F \wedge F \wedge F \wedge F
\right],
\end{align}
where we have defined
\begin{align}
(F \wedge F \pm *_8 F \wedge F)^2
=
(F \wedge F \pm *_8 F \wedge F) \wedge *_8
(F \wedge F \pm *_8 F \wedge F).
\end{align}
Then the action is bounded from below by the fourth Chern number
$S \ge \pm \frac{\alpha}{\kappa g^2} \int \! \mathrm{Tr} [F \wedge F \wedge F \wedge F]$.
The Bogomol'nyi bound is saturated when \eqref{eq:self-dual_R8_1} is satisfied.
It is easy to show that the (anti-)self-dual solution satisfies
the full equation of motion for the quartic model \eqref{eq:quartic}.

This kind of quartic Yang-Mills theory is not the standard gauge field
theory but appears in some physically appropriate situations.
For example, the quartic model
has been introduced to provide the
solution for the fundamental strings in $\text{SO}(32)$ heterotic string theory
\cite{Duff:1990wu, Duff:1991sz}. Tree-level heterotic five-brane is expected to
induce a quartic Yang-Mills theory whose (anti-)self-dual instantons in eight
dimensions precisely reproduce the energy-momentum tensor for
fundamental strings. On the other hand, the quadratic Yang-Mills part
is expected to appear at the one-loop level in perturbative heterotic
five-brane theory \cite{Duff:1990hb}.
This is in contrast to the heterotic fundamental string theory where the
quartic Yang-Mills term appears in the one-loop level.

There are other applications of the quartic model.
For example, various topological solitons specific for the quartic model
have been studied in \cite{Tchrakian:1978sf, Bernevig:2003yz, Kihara:2004yz}.

\paragraph{D7-brane effective action and D-instantons}
We next consider more physically relevant models.
Higher dimensional gauge theories are naturally realized as low-energy effective
field theories on D-branes.
The $(p+1)$-dimensional quadratic Yang-Mills theory appears in the
zero-slope limit $\alpha' \to 0$ of the open string sector on
D$p$-branes.
The four-dimensional instantons are interpreted as D-instantons (or
D$(-1)$-branes) embedded in D3-branes \cite{Douglas:1995bn}.
The ADHM moduli are interpreted as the zero-dimensional fields on the
D-instanton world-volume.
The ADHM constraint comes from the supersymmetric D-term condition of the
D3-D$(-1)$, D$(-1)$-D$(-1)$ open string sectors \cite{Witten:1995gx}.
This interpretation is generalized to D$p$-D$(p-4)$ brane systems.

For the eight-dimensional gauge theory, we consider Euclidean D7-branes in type IIB
string theory.
In order to see the (anti-)self-dual instanton effects, we consider the
$\alpha'$ corrections to the eight-dimensional Yang-Mills theory.
This is obtained from the dimensional reduction of the $\alpha'$
corrected super Yang-Mills theory in ten-dimensions \cite{Bergshoeff:2001dc}.
The gauge field part of the D7-brane world-volume Lagrangian is given by
\begin{align}
\mathcal{L}_{F} =
\frac{1}{\kappa g^2} \mathrm{Tr}
\left[
- \frac{1}{4} F_{\mu \nu} F^{\mu \nu}
\right]
+ \frac{
(2\pi \alpha')^2
}{12 \kappa g^2}
t_8 \mathrm{Tr} F^4 + \mathcal{O} \left( \frac{
\alpha^{\prime 4}
}{g^2} \right).
\label{eq:D7wv}
\end{align}
Here the first term is the eight-dimensional quadratic Yang-Mills part.
The second part is the first $\alpha'$ correction given by
\begin{align}
t_8 \mathrm{Tr} F^4 =& \
- 4 \mathrm{Tr}
\left[
\frac{1}{32} F_{\mu \nu} F_{\rho \sigma} F^{\mu \nu} F^{\rho \sigma}
+
\frac{1}{16} (F_{\mu \nu} F^{\mu \nu})^2
\right.
\notag \\
& \
\qquad \qquad
\left.
-
\frac{1}{8} F_{\mu \nu} F^{\nu \rho} F_{\rho \sigma} F^{\sigma \mu}
-
\frac{1}{4} F^{\mu \rho} F_{\rho \nu} F_{\mu \sigma} F^{\sigma \nu}
\right].
\end{align}
We now interpret the eight-dimensional instantons as the D-instantons
embedded in the D7-branes.
The D-instantons are the sources of the R-R 0-form $C^{(0)}$.
The $C^{(0)}$ coupling to the D7-branes is given by the Wess-Zumino term
of the effective action:
\begin{align}
\mathcal{L}_{\mathrm{WZ}}
= \frac{\mu_7}{\kappa g^2} \mathrm{Tr}
\left[
\frac{
(2 \pi \alpha')^2
}{4!} C^{(0)} F \wedge F \wedge F \wedge F
\right]
= \frac{\mu_7}{\kappa g^2} \mathrm{Tr}
\left[
\frac{
(2\pi \alpha')^2
}{4! \cdot 2^4} C^{(0)} \varepsilon^{\mu_1 \cdots \mu_8} F_{\mu_1 \mu_2}
 \cdots F_{\mu_7 \mu_8}
\right] d^8x.
\end{align}
Here $\mu_7$ is the R-R charge of the D7-brane.
In order that the eight-dimensional instantons $F \wedge F = \pm *_8 F
\wedge F$ whose instanton number
$ k = \frac{1}{4! (2\pi)^4}  \mathrm{Tr} \int F \wedge F \wedge F \wedge F$
become the source of the R-R 0-form,
the quartic term in $\mathcal{L}_F$ evaluated on the instantons
should coincide with $\mathcal{L}_{\text{WZ}}$.
We also need the condition that the quadratic term in $\mathcal{L}_F$ on
the instantons vanish.
Therefore the eight-dimensional (anti-)self-dual instantons become the
D-instantons when the following conditions are satisfied:
\begin{align}
\frac{1}{g^2}
\int \! d^8 x \ \mathrm{Tr}
[F_{\mu\nu} F^{\mu\nu}]
= 0, \qquad
\frac{1}{12} t_8 F^4 = \ \pm \frac{1}{4! \cdot 2^4} \varepsilon^{\mu_1 \cdots \mu_8} F_{\mu_1 \mu_2}
 \cdots F_{\mu_7 \mu_8}, \label{Dinst_cond}
\end{align}
and all the $\mathcal{O}(\alpha^{\prime 4}/g^2)$ terms vanish.
We call \eqref{Dinst_cond} the D-instanton conditions.
When the D-instanton conditions holds on the instanton solution, then,
the quartic term in $\mathcal{L}_F$ agrees with
the effective action of $k$ D-instantons:
\begin{align}
S_{D(-1)} = \mu_{-1} C^{(0)} \frac{1}{\kappa} \mathrm{Tr} \mathbf{1}_k =
 \frac{k \mu_{-1}}{\kappa} C^{(0)}.
\label{eq:D-inst_wv}
\end{align}
Here $\mu_{-1}$ is the R-R charge of the D-instanton 
and we have used the relation $\frac{\lambda^2}{g^2_7} =
\frac{1}{(2\pi)^3 g_s}$, $\mu_{-1} = \frac{2\pi}{g_s}$ and $g_s$ is the
string coupling.

We consider the zero-slope limit $\alpha' \to 0$ with fixed
$\alpha^{\prime 2}/g^2$
 to obtain the effective action \eqref{eq:D7wv}.
In this limit, the $\mathcal{O}(\alpha^{\prime 4}/g^2)$ terms vanish and
the $F$ quartic term remains
finite while the Yang-Mills part diverges in general.
The situation where the conditions \eqref{Dinst_cond} are satisfied
has been analyzed in \cite{Billo:2009di} where
the instanton partition function for the D7/D$(-1)$ system is studied.
In there, it is shown that the one-instanton solution
\cite{Grossman:1984pi, Tchrakian:1984gq}
satisfies the D-instanton condition \eqref{Dinst_cond} when the size
modulus becomes zero.

We find that our general solution \eqref{eq:R8_ADHM_gauge_zero_mode}
actually satisfies the second condition in \eqref{Dinst_cond}.
This is due to the property of the basis $e_{\mu}$ defined by the
Clifford algebra.
For the first condition in \eqref{Dinst_cond},
we can evaluate the quadratic Yang-Mills term for the 't Hooft type
instantons in the
well-separated limit \eqref{eq;R8_instantons_well-separated_limit}.
The Yang-Mills quadratic term becomes
\begin{align}
\frac{1}{g^2}
\int \! d^8 x \ \mathrm{Tr} [F_{\mu \nu} F^{\mu \nu}] =& \
\frac{1}{g^2}
\mathrm{Tr} [\Sigma^{(-)}_{\mu \nu} \Sigma^{(-) \mu \nu}]
\sum_{p=1}^k
\int \! d^8 x \
\left(
\frac{\lambda_p^2}{(\lambda_p^2 + \|\tilde{x}_p \|^2)^2}
\right)^2.
\label{eq:Dinst1}
\end{align}
This is just the $k$ times the one-instanton contribution.
The radial part of the space-time integral in \eqref{eq:Dinst1} has been
calculated to be \cite{Billo:2009gc,Billo:2009di}
\begin{align}
\int^{\Lambda}_0 \! r^7 dr
\left(
\frac{\lambda_p^2}{(\lambda_p^2 + r^2)^2}
\right)^2 =
\frac{\lambda_p^2}{2} \log
\left(
1 + \frac{\Lambda^2}{\lambda_p^2}
\right) - \frac{11}{12} \lambda_p^2,
\end{align}
where we have  
introduced the cutoff $\Lambda$ in the space-time integral
and neglected sub-leading terms of $1/\Lambda$.
Then, when the instantons shrink to zero-size $\lambda_p \to 0$, the first condition in
\eqref{Dinst_cond} is satisfied
\footnote{
This zero-size limit should be taken
so that $\frac{\lambda_p^2}{g^2} \log (\frac{\Lambda_p^2}{\lambda_p^2}) \to
0$ in the limit $\alpha' \to 0$ with fixed $\alpha^{\prime 2}/g^2$.
}.
This result is a multi-instanton generalization of the one-instanton
calculations in \cite{Billo:2009gc,Billo:2009di}.
Therefore we conclude that the (anti-)self-dual instantons in the small instanton
limit correspond to the D-instantons embedded in the D7-branes.
We emphasize that the small instanton $\lambda_p \to 0$ is the strict
limit of the dilute instanton gas approximation
\eqref{eq;R8_instantons_well-separated_limit}.
In this limit,
all the instantons show singular behavior and they satisfy the equation
of motion.
Note that the fourth Chern number is kept finite in this limit.
The string origin of the zero-size limit of instantons is also discussed
in \cite{Minasian:2001ib}.

A few comments are in order.
First, the $\text{SO}(8)$ instantons in eight dimensions are studied in the context of
hetrotic/type I string duality \cite{Polchinski:1995df}.
Consider the (Euclidean) D7-branes in type IIB orientifold theory compactified on
two torus $T^2$.
The D7-branes are placed on top of the O7-planes.
There are four $\text{SO}(8)$ sectors in the theory.
Let us concentrate on the one sector among them.
The world-volume theory of the D7-branes is given by the eight-dimensional $\mathcal{N} = 2$ super
Yang-Mills theory with the gauge group $\text{SO}(8)$.
The self-dual instantons give non-perturbative effects in
eight-dimensional gauge theories \cite{Fucito:2009rs, Billo:2011uc}.
This is a non-perturbative test of the string duality.

Second,
the famous anomaly cancellation term $B \mathrm{Tr} F^4$,
where $B$ is the NS-NS B-field,
in heterotic string theory indicates that configurations with the finite
fourth Chern number become sources of fundamental strings
\cite{Green:1984sg}.
This configuration is nothing but the (anti-)self-dual instanton in eight dimensions.

\section{Conclusion and discussions}
In this paper we have studied ADHM construction of (anti-)self-dual instantons in eight
dimensions. The instantons satisfy the (anti-)self-dual equations $F \wedge
F = \pm *_8 F \wedge F$.
The gauge field is given by the pure gauge form \eqref{eq:R8_ADHM_gauge_zero_mode} which is a natural
generalization of the four-dimensional (anti-)self-dual instantons.
The ADHM construction is based on the basis $e_{\mu}$ which is
constructed from the Clifford algebra in seven dimensions.
Due to the property of the basis $e_{\mu}$, the eight-dimensional anti-self-dual equation reduces
to a set of algebraic constraints on matrices (the ADHM data).
Compared with the (anti-)self-dual equation in four dimensions, the equation in
eight dimensions is non-linear in $F$ and contains terms with space-time
derivative of second order.
We have found that there are the first and the second ADHM constraints on the
ADHM data. The former is the same form of the four-dimensional one 
while the latter comes from the non-linearity of the equation
and essentially a new ingredient.
We have also pointed out that the gauge group of the theory is determined
by the structure of the seven-dimensional Clifford algebra.

We have shown that our construction precisely
reproduces the known one-instanton profile, namely, the $\text{SO}(8)$
instanton \cite{Grossman:1984pi, Tchrakian:1984gq}.
We have also found the $k=2,3$
multi-instanton solutions based on the 't Hooft and JNR ansatz.
The JNR type solution contain more moduli parameters compared with the 't Hooft type.
We have shown that 
the first and the second ADHM constraints are explicitly solved in the dilute
instanton gas approximation.
The topological charges are evaluated numerically and we have shown that
the consistent results are found in a good accuracy.
It is obvious that any higher charge
solutions can be systematically constructed.
Although they are approximate solutions,
as far as we know, they are the first explicit examples of higher charge
solutions that do not show spherical symmetry in eight dimensions.

We have discussed the eight-dimensional gauge theories where the
(anti-)self-dual equation is relevant.
The instanton configurations extremize the action of the quadratic field
strength. Therefore the theory inevitably contain higher derivative terms.
As in the four-dimensional case, the eight-dimensional ADHM construction
enjoys the space-time gauge symmetry and the gauge symmetry in the instanton
space (dual space). This fact strongly suggests that the ADHM construction presented
in this paper has string theory origin in D-brane configurations
\cite{Witten:1995gx, Douglas:1996uz}.
Indeed, in \cite{Billo:2009gc,Billo:2009di}, the authors studied D7/D$(-1)$-brane
configurations in type IIB orientifold.
The open string scattering amplitudes including zero-modes associated
with strings that end on these branes reveal that the moduli action for
eight-dimensional $k=1$ self-dual instanton is given by the D-instanton
effective action.
We have 
exhibited a strong evidence
that this is true even for the multi-instantons in the small instanton limit.
In this limit, the 't Hooft type multi-instantons become exact
solutions of the (anti-)self-dual equation.
They also satisfy the D-instanton conditions in this limit and
identified with the D-instantons embedded in the D7-brane world-volume.

In four dimensions, the ADHM construction of instantons in noncommutative
space has been studied where the ADHM constraint is modified by the
noncommutativity parameter.
It is interesting to study the ADHM construction of instantons in
noncommutative space-time in eight dimensions.
It is also interesting to study monopoles in seven dimensions
\cite{Burzlaff:1986gv} and its Nahm construction.
Using the ADHM data we can also construct calorons in seven dimensions.
In the high temperature limit, we expect that the seven-dimensional
monopoles are realized.

It is interesting to study (anti-)self-dual equations in dimensions greater than eight.
In $4n$ dimensions, we can consider the (anti-)self-dual equation
$F \wedge F \wedge \cdots \wedge F = \pm *_{4n} F \wedge F \wedge \cdots
 \wedge F$,
where $F \wedge F \wedge \cdots \wedge F$ in both sides are $2n$-forms.
Solutions to the (anti-)self-dual equation are expected to have finite $2n$-th Chern number
$k = \frac{1}{(2n)!(2\pi)^{2n}} \int \! \mathrm{Tr} [F \wedge F \wedge \cdots
\wedge F],$ for appropriate basis $e_{\mu}$.
We find that the ADHM construction of instantons in eight dimensions
presented in this paper is generalized to $4n$ dimensions \cite{NaSaTa}.
Supersymmetric generalization including higher derivative interactions 
\cite{Nitta:2014pwa} is also important to study the relation to string
theories.
We will come back to these issues in future studies.

%%%%%%%%%%%%%%
\subsection*{Acknowledgments}
We would like to thank Y.~Amari, T.~Fujimori, M.~Hamanaka, H.~Nakajima,
M.~Nitta, S.~Terashima and Y.~Yasui for useful discussions and comments.
The work of S. S. is supported in part by Kitasato University
Research Grant for Young Researchers.
%%%%%%%%%%%%%%%%%%%%%%%%%%%%%%%%%%%

\begin{appendix}

\section{ADHM construction with U$(2)$ gauge in four dimensions} \label{sec:4-dim_ADHM_U(2)_gauge}
In this section, we give a brief review on the ADHM construction of instantons in four dimensions. We consider the gauge group U$(2)$.

The four-dimensional Weyl operator $\Delta_{(4)}$ is defined by
\begin{equation}
\Delta_{(4)} = C(x\otimes\mathbf{1}_k)+D,	\label{eq:Weyl_operator_R4}
\end{equation}
where $C$ and $D$ are quaternionic $(k+1)\times k$ matrices, $k$ is the instanton charge 
and $x = x^{\mu}e_{\mu}$.
Here $x^{\mu}~(\mu=1,\dots,4)$ is the Cartesian coordinate of the
four-dimensional Euclid space, $e_{\mu}=(-i\sigma_i,\mathbf{1}_2)$ is
the basis of the quaternion and $\sigma_i$ are the Pauli matrices.
The Weyl operator $\Delta_{(4)}$ is assumed to satisfy the ADHM constraint:
\begin{equation}
\Delta^{\dagger}_{(4)}\Delta_{(4)} = \mathbf{1}_2\otimes E_k,	\label{eq:ADHM_4-dim}
\end{equation}
where $\Delta^{\dagger}_{(4)}$ is the quaternionic conjugate of $\Delta_{(4)}$ and $E_k$ is an invertible $k\times k$ matrix.

In order to construct the instanton solution for the gauge field $A_{\mu}(x)$, it is necessary to find a quaternionic $(k+1)$ column vector $V(x)$ obeying the Weyl equation:
\begin{equation}
\Delta^{\dagger}_{(4)}V(x) = 0,
\end{equation}
where $V(x)$ is the zero-mode normalized as $V^{\dagger}(x)V(x)=
\mathbf{1}_2$.
The gauge field $A_{\mu}(x)$ of instantons is given by
\begin{equation}
A_{\mu}(x) = V^{\dagger}(x)\partial_{\mu}V(x).	\label{eq:ADHM_gauge_from_zero_modes}
\end{equation}
Using the expression \eqref{eq:ADHM_gauge_from_zero_modes}, the field strength is calculated as
\begin{align}
F_{\mu\nu} &= \partial_{\mu}V^{\dagger}\left( 
\mathbf{1}_{2+2k}
-VV^{\dagger} \right)\partial_{\nu}V - (\mu \leftrightarrow \nu).	\label{eq:ADHM_F_zero_mode_expand_1}
\end{align}
Here we use the completeness relation:
\begin{equation}
\mathbf{1}_{2+2k}-VV^{\dagger} = \Delta_{(4)}(\Delta^{\dagger}_{(4)}\Delta_{(4)})^{-1}\Delta^{\dagger}_{(4)}.	\label{eq:Weyl_operater_completeness_relation}
\end{equation}
Then \eqref{eq:ADHM_F_zero_mode_expand_1} is rewritten as
\begin{align}
F_{\mu\nu} &= V^{\dagger}C(e_{\mu}\otimes\mathbf{1}_k)(\Delta_{(4)}^{\dagger}\Delta_{(4)})^{-1}(e_{\nu}^{\dagger}\otimes\mathbf{1}_k)C^{\dagger}V - (\mu \leftrightarrow \nu) \notag \\
&= V^{\dagger}C(\Delta_{(4)}^{\dagger}\Delta_{(4)})^{-1} \left( \eta^{(-)}_{\mu\nu}\otimes\mathbf{1}_k\right)C^{\dagger}V,
\end{align}
where we have used the ADHM constraint \eqref{eq:ADHM_4-dim}. Here $\eta^{(\pm)}_{\mu\nu}$ is the 't Hooft symbol defined by
\begin{equation}
\eta^{(+)}_{\mu\nu} = e^{\dagger}_{\mu}e_{\nu}-e^{\dagger}_{\nu}e_{\mu},~~~
\eta^{(-)}_{\mu\nu} = e_{\mu}e^{\dagger}_{\nu}-e_{\nu}e^{\dagger}_{\mu}.	\label{eq:'tHooft_tensor}
\end{equation}
The 't Hooft symbol satisfies the four-dimensional (anti-)self-dual relation:
\begin{equation}
\eta^{(\pm)}_{\mu\nu} = \pm\frac{1}{2!}\varepsilon_{\mu\nu\rho\sigma}\eta^{(\pm)}_{\rho\sigma}.
\end{equation}
Therefore the field strength $F_{\mu\nu}$ associated with the solution \eqref{eq:ADHM_gauge_from_zero_modes} automatically satisfies the (anti-)self-dual equation $F=\pm\ast_4 F$.

From the above discussion, we find that a key point of the ADHM construction is that the 't Hooft symbol $\eta^{(\pm)}_{\mu\nu}$ constructed from the basis $e_{\mu}$ satisfies the (anti-)self-dual relation.
Therefore if we formulate the ADHM construction of instantons in higher dimensions then we need to find the basis that satisfies the (anti-)self-dual relation in higher dimensions.

\section{Clifford algebra and $4n$-dimensional (anti-)self-dual tensor}	\label{sec:Clifford_ASD-tensor}
In this section, we construct the $4n$-dimensional generalization of the 't Hooft symbol which satisfies the (anti-)self-dual relation.
We first introduce $(m=4n-1)$-dimensional Clifford algebra $C\ell_m(K)$ on the (number) field $K$. Elements of the Clifford algebra $\Gamma_i\in C\ell_m(K)$ satisfy the relation:
\begin{equation}
\Gamma_i\Gamma_j+\Gamma_j\Gamma_i=-2\delta_{ij},
\end{equation}
where the indices $i,j$ run from $1$ to $4n-1$.
For $K=\mathbb{R},$ $C\ell_m(\mathbb{R})$ is called ``the real Clifford algebra''. On the other hand, for $K=\mathbb{C},$ $C\ell_m(\mathbb{C})$ is called ``the complex Clifford algebra''.
In $4n-1$ dimensions, the chirality element $\omega$ is defined by
\begin{subequations}
\begin{align}
&\omega= (-1)^{\left\lfloor (m+5)/4 \right\rfloor}\Gamma_1\Gamma_2\dots \Gamma_m,~~~\Gamma_i\in C\ell_m(\mathbb{R}),  \label{eq:real_Clifford_chiral_element} \\
&\omega= i^{\left\lfloor (m+5)/2 \right\rfloor}\Gamma_1\Gamma_2\dots \Gamma_m,~~~\Gamma_i\in C\ell_m(\mathbb{C}),	\label{eq:complex_Clifford_chiral_element}
\end{align}
\end{subequations}
where the symbol $\lfloor x \rfloor$ is the floor function (for example: $\lfloor 2.8 \rfloor=2,\lfloor 3 \rfloor=3$).
Here we define the overall factor of the chirality element $\omega$ for later convenience.
It is well known that we can decompose the $(4n-1)$-dimensional Clifford algebra by using the chirality element.
The projection operator is defined by
\begin{equation}
P_{\pm} = \frac{1}{2} ( 1\pm \omega ).
\end{equation}
Using $P_{\pm}$, we can decompose the Clifford algebra as
\begin{equation}
C\ell_m(K)=C\ell^{(+)}_m(K)\oplus C\ell^{(-)}_m(K),
\end{equation}
where $C\ell^{(\pm)}_m(K)$ are defined by elements in 
$C\ell_m (K)$ projected by $P_{\pm}$. We call $C\ell^{(\pm)}_m(K)$ ``the decomposed Clifford algebra''.
Now we choose the elements of the decomposed Clifford algebra
$\Gamma_i^{(\pm)}\in C\ell^{(\pm)}_m(K)$ that satisfy the relation $\Gamma_i^{(+)}=-\Gamma_i^{(-)}$.

Note that the elements of the decomposed Clifford algebra $\Gamma_i^{(\pm)}\in C\ell^{(\pm)}_m(K)$ satisfy the relation $\{ \Gamma_i^{(\pm)},\Gamma_j^{(\pm)} \}=-2\delta_{ij}$, but $\Gamma_i^{(\pm)}$ are not elements of the Clifford algebra.
Because the elements of the decomposed Clifford algebra are not the algebraic generators.
The algebraic generators have the property that each element of the algebra is not produced by a product of other elements, 
that is 
$e_ie_j\dots\neq e_t$
where
$e_i,e_j,\dots,e_t\in Q(K)$
 and $Q(K)$ is an algebra on the field $K$.
The elements of the Clifford algebra $\Gamma_i$ are algebraic
generators, therefore $\Gamma_i$ satisfies the relation
$\Gamma_i\Gamma_j \dots\neq\Gamma_t$
, where 
$\Gamma_i,\Gamma_j,\dots,\Gamma_t\in C\ell_m(K)$.
On the other hand, the element of the decomposed Clifford algebra $\Gamma_i^{(\pm)}$ does not satisfy
the relation 
$\Gamma_i^{(\pm)}\Gamma_j^{(\pm)} \dots \neq\Gamma_t^{(\pm)}$
, where 
$\Gamma_i^{(\pm)},\Gamma_j^{(\pm)},\dots,\Gamma_t^{(\pm)}\in
C\ell^{(\pm)}_m(K)$.

We can construct the $4n$-dimensional (anti-)self-dual tensor $\Sigma^{(\pm)}_{\mu\nu}$ form the $(4n-1)$-dimensional Clifford algebra $C\ell_{4n-1}(K)$.
Here the $4n$-dimensional ``(anti-)self-dual tensor'' means that the
tensor satisfies the (anti-)self-dual relation in $4n$ dimensions.
We define the $4n$-dimensional basis $e_{\mu}$ by
\begin{equation}
e_{\mu} = \delta_{\mu4n}1 + \delta_{\mu i}\Gamma^{(-)}_i,~~e^{\dagger}_{\mu} = \delta_{\mu4n}1 + \delta_{\mu i}\Gamma^{(+)}_i,	\label{eq:4n-dim_ASD_basis_from_Clifford}
\end{equation}
where $1$ is an identity element (such that $1\Gamma^{(\pm)}_i=\Gamma^{(\pm)}_i1$) and the indices $\mu,\nu,\dots$ run from $1$ to $4n$.
Using this basis, we define the $4n$-dimensional (anti-)self-dual tensor by
\begin{align}
\Sigma_{\mu\nu}^{(+)} = e^{\dagger}_{\mu}e_{\nu}-e^{\dagger}_{\nu}e_{\mu},~~~~
\Sigma_{\mu\nu}^{(-)} = e_{\mu}e^{\dagger}_{\nu}-e_{\nu}e^{\dagger}_{\mu}. 	\label{eq:ASD_tensor_from_pseudo-Clifford}
\end{align}
We can confirm that $\Sigma^{(\pm)}_{\mu\nu}$ satisfies the $4n$-dimensional (anti-)self-dual relation:
\begin{equation}
\Sigma^{(\pm)}_{[a_1a_2}\dots\Sigma^{(\pm)}_{a_{2n-1}a_{2n}]} = \pm\frac{1}{2n!}\varepsilon_{a_1a_2\dots a_{2n}b_1b_2\dots b_{2n}}\Sigma^{(\pm)}_{b_1b_2}\dots\Sigma^{(\pm)}_{b_{2n-1}b_{2n}}	\label{eq:4n_ASD_equations_ASD_tensor}
\end{equation}
where $\Sigma_{\mu\nu}^{(+)}$ satisfies the self-dual equation and $\Sigma_{\mu\nu}^{(-)}$ satisfies the anti-self-dual equation respectively.

In a $4n$-dimensional ADHM construction, we have to represent the $(4n-1)$-dimensional Clifford algebra $C\ell_{4n-1}(K)$ by matrices.
It is well known that the complex (real) Clifford algebra has an isomorphism with a matrix ring.
Furthermore the complex (real) Clifford algebra has the period with two
(eight) from the Bott periodicity theorem \cite{RauschdeTraubenberg:2005aa}.
Therefore we can naturally obtain the matrix representations of the complex (real) Clifford algebra (Table \ref{table:matrix-Clifford_algebra}).
\begin{table}[tbp]
\begin{center}
\begin{tabular}{c||c|c}
$4n$-dim. $\mod 8$ & $C\ell_{4n-1}(\mathbb{C})$ & $C\ell_{4n-1}(\mathbb{R})$ \\
\hline
$4$ & $\text{GL}(2^{2n-1};\mathbb{C})\oplus \text{GL}(2^{2n-1};\mathbb{C})$ & $\text{GL}(2^{2n-2};\mathbb{H})\oplus \text{GL}(2^{2n-2};\mathbb{H})$ \\
$8$ & $\text{GL}(2^{2n-1};\mathbb{C})\oplus \text{GL}(2^{2n-1};\mathbb{C})$ & $\text{GL}(2^{2n-1};\mathbb{R})\oplus \text{GL}(2^{2n-1};\mathbb{R})$
\end{tabular}
\caption{The matrix rings $\text{GL}(N;K)$ which are isomorphic to the $(4n-1)$-dimensional complex (real) Clifford algebra $C\ell_{4n-1}(\mathbb{C}(\mathbb{R}))$.
Here $N$ is the matrix size and the symbol $\mathbb{H}$ means the quaternion.}
\label{table:matrix-Clifford_algebra}
\end{center}
\end{table}

Note that the gauge group of the ADHM construction based on the
(anti-)self-dual tensor \eqref{eq:ASD_tensor_from_pseudo-Clifford} is
determined by the (number) field of the Clifford algebra.
Therefore the size of the gauge group (color size) $N$ is dependent on a matrix size of the matrix representation of the Clifford algebra $\text{GL}(N;K)$.

Now we have obtained the $4n$-dimensional (anti-)self-dual tensor.
We construct the four- and eight-dimensional (anti-)self-dual basis explicitly.
Note that the representation of the basis is not unique. We use the tensor product of the following $2\times2$ matrices.
The complex Clifford algebra $C\ell_m(\mathbb{C})$ is constructed by the Pauli matrices:
\begin{equation}
\sigma_1 = \begin{pmatrix} 0 & 1 \\ 1 & 0 \end{pmatrix},~~\sigma_2 = \begin{pmatrix} 0 & -i \\ i & 0 \end{pmatrix},~~\sigma_3 = \begin{pmatrix} 1 & 0 \\ 0 & -1 \end{pmatrix},~~~ \sigma_0=\mathbf{1}_2.
\end{equation}
On the other hand, the real Clifford algebras $C\ell_m(\mathbb{R})$ are constructed by the following matrices \cite{Lundell:1989ry}:
\begin{equation}
\tau_1 = \begin{pmatrix} 0 & 1 \\ 1 & 0 \end{pmatrix},~~\tau_2 = \begin{pmatrix} 0 & -1 \\ 1 & 0 \end{pmatrix},~~\tau_3 = \begin{pmatrix} 1 & 0 \\ 0 & -1 \end{pmatrix},~~~\tau_0=\mathbf{1}_2.
\end{equation}
For simplicity, we omit the tensor (Kronecker) product symbol $\otimes$ in the following discussions. For example, $\sigma_{ij}$ means $\sigma_i\otimes\sigma_j$.

\paragraph{The complex basis in four dimensions}
We construct the four-dimensional (anti-)self-dual tensor from the three-dimensional Clifford algebra.
The matrix representation of the three-dimensional complex Clifford algebra $C\ell_3(\mathbb{C})$ is given by
\begin{equation}
\Gamma_1 = \begin{pmatrix} i\sigma_1 & 0 \\ 0 & -i\sigma_1 \end{pmatrix},~~  \Gamma_2 = \begin{pmatrix} i\sigma_2 & 0 \\ 0 & -i\sigma_2 \end{pmatrix},~~  \Gamma_3 = \begin{pmatrix} i\sigma_3 & 0 \\ 0 & -i\sigma_3 \end{pmatrix}.
\end{equation}
The chiral matrix $\omega$ and the projection operators $P_{\pm}$ are
\begin{equation}
\omega = \Gamma_1\Gamma_2\Gamma_3 = \begin{pmatrix} \mathbf{1}_2 & 0 \\ 0 & -\mathbf{1}_2 \end{pmatrix},
\hspace{30pt}
P_{+} = \begin{pmatrix} \mathbf{1}_2 & 0 \\ 0 & 0 \end{pmatrix},~~  P_{-} = \begin{pmatrix} 0 & 0 \\ 0 & \mathbf{1}_2 \end{pmatrix}.
\end{equation}
Using these matrices, we obtain
\begin{equation}
\Gamma_i^{(\pm)} = \pm i\sigma_i,
\end{equation}
where $i=1,2,3$. Therefore we obtain the four-dimensional (anti-)self-dual complex basis:
\begin{equation}
e_{\mu} = \delta_{\mu4}\mathbf{1}_2 - i\delta_{\mu i}\sigma_i,~~e^{\dagger}_{\mu} = \delta_{\mu4}\mathbf{1}_2 + i\delta_{\mu i}\sigma_i.
\end{equation}
This basis is nothing but the quaternion basis which is used in the four-dimensional ADHM construction.
In the previous discussion in subsection \ref{subsec:gauge_homotopy_group}, the gauge group is U$(2)$ for this basis.

\paragraph{The real basis in four dimensions}
For Table \ref{table:matrix-Clifford_algebra}, the three-dimensional real Clifford algebra $C\ell_3(\mathbb{R})$ is isomorphic to $\mathbb{H}\oplus\mathbb{H}$. However we use real matrix representation to implement gauge group O$(4)$.
The real matrix representation of $C\ell_3(\mathbb{R})$ is given by
\begin{equation}
\Gamma_1 = \begin{pmatrix} \tau_{12} & 0 \\ 0 & -\tau_{12} \end{pmatrix},~~  \Gamma_2 = \begin{pmatrix} \tau_{20} & 0 \\ 0 & -\tau_{20} \end{pmatrix},~~  \Gamma_3 = \begin{pmatrix} \tau_{32} & 0 \\ 0 & -\tau_{32} \end{pmatrix}.
\end{equation}
The chiral matrix $\omega$ and the projection operators $P_{\pm}$ are
\begin{equation}
\omega = \Gamma_1\Gamma_2\Gamma_3 = \begin{pmatrix} -\mathbf{1}_4 & 0 \\ 0 & \mathbf{1}_4 \end{pmatrix},
\hspace{30pt}
P_{+} = \begin{pmatrix} 0 & 0 \\ 0 & \mathbf{1}_4 \end{pmatrix},~~  P_{-} = \begin{pmatrix} \mathbf{1}_4 & 0 \\ 0 & 0 \end{pmatrix}.
\end{equation}
Therefore $\Gamma_i^{(\pm)}$ are
\begin{equation}
\Gamma_1^{(\pm)} = \mp \tau_{12},~~~ \Gamma_2^{(\pm)} = \mp \tau_{20},~~~ \Gamma_3^{(\pm)} = \mp \tau_{32},
\end{equation}
and we obtain the four-dimensional (anti-)self-dual tensor by using \eqref{eq:4n-dim_ASD_basis_from_Clifford} and \eqref{eq:ASD_tensor_from_pseudo-Clifford}.
If this real basis is used in the four-dimensional ADHM construction, the gauge group becomes $G=\text{O}(4)$.

\paragraph{The complex basis in eight dimensions}
The matrix representation of $C\ell_7(\mathbb{C})$ is given by
\begin{align}
&\Gamma_1 = \begin{pmatrix} i\sigma_{133} & 0 \\ 0 & -i\sigma_{133} \end{pmatrix},~  \Gamma_2 = \begin{pmatrix} i\sigma_{233} & 0 \\ 0 & -i\sigma_{233} \end{pmatrix},~  \Gamma_3 = \begin{pmatrix} i\sigma_{013} & 0 \\ 0 & -i\sigma_{013} \end{pmatrix},  \notag \\
&\Gamma_4 = \begin{pmatrix} i\sigma_{023} & 0 \\ 0 & -i\sigma_{023} \end{pmatrix},~  \Gamma_5 = \begin{pmatrix} i\sigma_{001} & 0 \\ 0 & -i\sigma_{001} \end{pmatrix},~  \Gamma_6 = \begin{pmatrix} i\sigma_{002} & 0 \\ 0 & -i\sigma_{002} \end{pmatrix},~ \Gamma_7 = \begin{pmatrix} i\sigma_{333} & 0 \\ 0 & -i\sigma_{333} \end{pmatrix}.
\end{align}
Using \eqref{eq:complex_Clifford_chiral_element}, the chiral matrix $\omega$ is given by
\begin{equation}
\omega = (-1)\Gamma_1\Gamma_2\Gamma_3\Gamma_4\Gamma_5\Gamma_6\Gamma_7 = \begin{pmatrix} \mathbf{1}_8 & 0 \\ 0 & -\mathbf{1}_8 \end{pmatrix}.
\end{equation}
The projection operators $P_{\pm}$ are
\begin{equation}
P_{+} = \begin{pmatrix} \mathbf{1}_8 & 0 \\ 0 & 0 \end{pmatrix},~~  P_{-} = \begin{pmatrix} 0 & 0 \\ 0 & \mathbf{1}_8 \end{pmatrix}.
\end{equation}
Therefore we obtain
\begin{align}
&\Gamma_{1}^{(\pm)} = \pm i\sigma_{133},~~\Gamma_{2}^{(\pm)} = \pm i\sigma_{233},~~\Gamma_{3}^{(\pm)} = \pm i\sigma_{013},\notag \\
&\Gamma_{4}^{(\pm)} = \pm i\sigma_{023},~~\Gamma_{5}^{(\pm)}=\pm i\sigma_{001},~~\Gamma_{6}^{(\pm)}=\pm i\sigma_{002},~~\Gamma_{7}^{(\pm)}=\pm i\sigma_{333}.
\label{eq:R8_complex_basis_I}
\end{align}
Of course, we can take another matrix representation:
\begin{align}
&\Gamma_1 = \begin{pmatrix} i\sigma_{112} & 0 \\ 0 & -i\sigma_{112} \end{pmatrix},~  \Gamma_2 = \begin{pmatrix} i\sigma_{120} & 0 \\ 0 & -i\sigma_{120} \end{pmatrix},~  \Gamma_3 = \begin{pmatrix} -i\sigma_{132} & 0 \\ 0 & i\sigma_{132} \end{pmatrix},  \notag \\
&\Gamma_4 = \begin{pmatrix} -i\sigma_{221} & 0 \\ 0 & i\sigma_{221} \end{pmatrix},~  \Gamma_5 = \begin{pmatrix} i\sigma_{223} & 0 \\ 0 & -i\sigma_{223} \end{pmatrix},~  \Gamma_6 = \begin{pmatrix} -i\sigma_{202} & 0 \\ 0 & i\sigma_{202} \end{pmatrix},~ \Gamma_7 = \begin{pmatrix} i\sigma_{300} & 0 \\ 0 & -i\sigma_{300} \end{pmatrix}.
\end{align}
In this case, $\Gamma^{(\pm)}_i$ are
\begin{align}
&\Gamma_1^{(\pm)} = \pm i\sigma_{112},~~ \Gamma_2^{(\pm)} = \pm i\sigma_{120},~~ \Gamma_3^{(\pm)} =
 \mp i\sigma_{132},\notag \\
&\Gamma_4^{(\pm)} = \mp i\sigma_{221},~~ \Gamma_5^{(\pm)} = \pm i\sigma_{223},~~ \Gamma_6^{(\pm)} = \mp i\sigma_{202},~~~~ \Gamma_7^{(\pm)} =
\pm i\sigma_{300},
\label{eq:R8_complex_basis_II}
\end{align}
The basis \eqref{eq:R8_complex_basis_II} is used to construct the Grossman's one-instantons \cite{Grossman:1984pi}.
These bases take complex values and the matrix size of $\Gamma_i^{(\pm)}$ is eight. 
Therefore the gauge group becomes U$(8)$ for this basis.

\paragraph{The real basis in eight dimensions}
The matrix representation of $C\ell_7(\mathbb{R})$ is given by
\begin{align}
&\Gamma_1 = \begin{pmatrix} \tau_{222} & 0 \\ 0 & -\tau_{222} \end{pmatrix},~~  \Gamma_2 = \begin{pmatrix} \tau_{012} & 0 \\ 0 & -\tau_{012} \end{pmatrix},~~  \Gamma_3 = \begin{pmatrix} \tau_{201} & 0 \\ 0 & -\tau_{201} \end{pmatrix},  \notag \\
&\Gamma_4 = \begin{pmatrix} \tau_{032} & 0 \\ 0 & -\tau_{032} \end{pmatrix},~~  \Gamma_5 = \begin{pmatrix} \tau_{120} & 0 \\ 0 & -\tau_{120} \end{pmatrix},~~  \Gamma_6 = \begin{pmatrix} \tau_{320} & 0 \\ 0 & -\tau_{320} \end{pmatrix},~~ \Gamma_7 = \begin{pmatrix} \tau_{203} & 0 \\ 0 & -\tau_{203} \end{pmatrix}.
\end{align}
Using \eqref{eq:real_Clifford_chiral_element}, the chiral matrix $\omega$ is given by
\begin{equation}
\omega = (-1)\Gamma_1\Gamma_2\Gamma_3\Gamma_4\Gamma_5\Gamma_6\Gamma_7 = \begin{pmatrix} \mathbf{1}_8 & 0 \\ 0 & -\mathbf{1}_8 \end{pmatrix}.
\end{equation}
The projection operators $P_{\pm}$ are
\begin{equation}
P_{+} = \begin{pmatrix} \mathbf{1}_8 & 0 \\ 0 & 0 \end{pmatrix},~~  P_{-} = \begin{pmatrix}  0 & 0 \\ 0 & \mathbf{1}_8 \end{pmatrix}.
\end{equation}
Therefore we obtain
\begin{align}
&\Gamma_{1}^{(\pm)} = \pm \tau_{222},~~\Gamma_{2}^{(\pm)} = \pm \tau_{012},~~\Gamma_{3}^{(\pm)} = \pm \tau_{201},\notag \\
&\Gamma_{4}^{(\pm)} = \pm \tau_{032},~~\Gamma_{5}^{(\pm)}=\pm \tau_{120},~~\Gamma_{6}^{(\pm)}=\pm \tau_{320},~~\Gamma_{7}^{(\pm)}=\pm \tau_{203}.
\label{eq:R8_real_basis}
\end{align}
This basis is real valued, therefore the gauge group becomes O$(8)$.

\section{Eight-dimensional U$(8)$ ADHM equations}	\label{sec:R8_U8-ADHM-eq}
In this section, we explicitly write down the eight-dimensional ADHM equations for U$(8)$ gauge group.
Here we use the complex basis \eqref{eq:R8_complex_basis_II}
\footnote{
Note that, we can use other basis \eqref{eq:R8_complex_basis_I}.
}
.
If we use the real basis \eqref{eq:R8_real_basis} then we obtain the eight-dimensional ADHM equations for O$(8)$ gauge group.

We assume that ADHM data $S$ is expanded by the basis $e_{\mu}$, that is
$S=e_{\mu}\otimes\tilde{S}^{\mu}$. The first ADHM equations are given by the following equations \eqref{eq:R8_ADHM-eq_cf_rr_1}, \eqref{eq:R8_ADHM-eq_cf_rr_2} and \eqref{eq:R8_ADHM-eq_cf_rr_3_4}.
\begin{align}
[T^2,T^5] - [T^3,T^6] + \frac{i}{2}\left( S^{\dagger}_4S_4 - S^{\dagger}_1S_1 \right) &=0,	\notag \\
[T^3,T^6] - [T^1,T^4] + \frac{i}{2}\left( S^{\dagger}_4S_4 - S^{\dagger}_2S_2 \right) &=0,	\notag \\
[T^1,T^4] - [T^2,T^5] + \frac{i}{2}\left( S^{\dagger}_4S_4 - S^{\dagger}_3S_3 \right) &=0,
\label{eq:R8_ADHM-eq_cf_rr_1}
\end{align}
\begin{align}
[T^1,T^4] + [T^2,T^5] - [T^3,T^6] - [T^8,T^7] - \frac{i}{2}\left( S^{\dagger}_3S_3-S^{\dagger}_7S_7 \right)  &= 0,	\notag \\
-[T^1,T^4] - [T^2,T^5] - [T^3,T^6] - [T^8,T^7] - \frac{i}{2}\left( S^{\dagger}_4S_4-S^{\dagger}_8S_8 \right) &= 0,	\label{eq:R8_ADHM-eq_cf_rr_3_4}
\end{align}
\begin{align}
[T^1,T^2] + [T^4,T^5] + \frac{1}{2}\left( S^{\dagger}_1S_2-S^{\dagger}_2S_1 \right) &=0,	&	[T^1,T^5] - [T^4,T^2] + \frac{i}{2}\left( S^{\dagger}_1S_2+S^{\dagger}_2S_1 \right) &=0,	\notag \\
[T^1,T^3] + [T^4,T^6] + \frac{1}{2}\left( S^{\dagger}_1S_3-S^{\dagger}_3S_1 \right) &=0,	&	[T^1,T^6] - [T^4,T^3] + \frac{i}{2}\left( S^{\dagger}_1S_3+S^{\dagger}_3S_1 \right) &=0,	\notag \\
[T^2,T^3] + [T^5,T^6] + \frac{1}{2}\left( S^{\dagger}_2S_3-S^{\dagger}_3S_2 \right) &=0,	&	[T^2,T^6] - [T^5,T^3] + \frac{i}{2}\left( S^{\dagger}_2S_3+S^{\dagger}_3S_2 \right) &=0,	\notag \\
[T^1,T^2] - [T^4,T^5] + \frac{1}{2}\left( S^{\dagger}_4S_3-S^{\dagger}_3S_4 \right) &=0,	&	[T^1,T^5] + [T^4,T^2] - \frac{i}{2}\left( S^{\dagger}_4S_3+S^{\dagger}_3S_4 \right) &=0,	\notag \\
[T^2,T^3] - [T^5,T^6] + \frac{1}{2}\left( S^{\dagger}_4S_1-S^{\dagger}_1S_4 \right) &=0,	&	[T^2,T^6] + [T^5,T^3] - \frac{i}{2}\left( S^{\dagger}_4S_1+S^{\dagger}_1S_4 \right) &=0,	\notag \\
[T^3,T^1] - [T^6,T^4] + \frac{1}{2}\left( S^{\dagger}_4S_2-S^{\dagger}_2S_4 \right) &=0,	&	[T^3,T^4] + [T^6,T^1] - \frac{i}{2}\left( S^{\dagger}_4S_2+S^{\dagger}_2S_4 \right) &=0,	\notag \\
[T^8,T^1] + [T^7,T^4] + \frac{1}{2}\left( S^{\dagger}_2S_7-S^{\dagger}_7S_2 \right) &=0,	&	[T^8,T^4] - [T^7,T^1] + \frac{i}{2}\left( S^{\dagger}_2S_7+S^{\dagger}_7S_2 \right) &=0,	\notag \\
[T^8,T^2] + [T^7,T^5] + \frac{1}{2}\left( S^{\dagger}_3S_5-S^{\dagger}_5S_3 \right) &=0,	&	[T^8,T^5] - [T^7,T^2] + \frac{i}{2}\left( S^{\dagger}_3S_5+S^{\dagger}_5S_3 \right) &=0,	\notag \\
[T^8,T^3] + [T^7,T^6] + \frac{1}{2}\left( S^{\dagger}_1S_6-S^{\dagger}_6S_1 \right) &=0,	&	[T^8,T^6] - [T^7,T^3] + \frac{i}{2}\left( S^{\dagger}_1S_6+S^{\dagger}_6S_1 \right) &=0,	\notag \\
[T^8,T^1] - [T^7,T^4] + \frac{1}{2}\left( S^{\dagger}_4S_5-S^{\dagger}_5S_4 \right) &=0,	&	[T^8,T^4] + [T^7,T^1] - \frac{i}{2}\left( S^{\dagger}_4S_5+S^{\dagger}_5S_4 \right) &=0,	\notag \\
[T^8,T^2] - [T^7,T^5] + \frac{1}{2}\left( S^{\dagger}_4S_6-S^{\dagger}_6S_4 \right) &=0,	&	[T^8,T^5] + [T^7,T^2] - \frac{i}{2}\left( S^{\dagger}_4S_6+S^{\dagger}_6S_4 \right) &=0,	\notag \\
[T^8,T^3] - [T^7,T^6] + \frac{1}{2}\left( S^{\dagger}_4S_7-S^{\dagger}_7S_4 \right) &=0,	&	[T^8,T^6] + [T^7,T^3] - \frac{i}{2}\left( S^{\dagger}_4S_7+S^{\dagger}_7S_4 \right) &=0.
\label{eq:R8_ADHM-eq_cf_rr_2}
\end{align}

Now an invertible $k\times k$ matrix $f$ is defined by the first ADHM constraint $\Delta^{\dagger}\Delta=\mathbf{1}_8\otimes f^{-1}$ (that is $f=(E_k^{(1)})^{-1}$).
The second ADHM equations are given by the following equations \eqref{eq:R8_additional_ADHM-eq_I}, \eqref{eq:R8_additional_ADHM-eq_II}.
\begin{align}
T^{\mu}f = fT^{\mu}	\label{eq:R8_additional_ADHM-eq_I}
\end{align}
\begin{align}
[T^1,T^2]&=0, & [T^2,T^3]&=0, & [T^3,T^1]&=0, & [T^4,T^5]&=0, & [T^5,T^6]&=0, & [T^6,T^4]&=0, \notag \\
[T^1,T^8]&=0, & [T^2,T^8]&=0, & [T^3,T^8]&=0, & [T^4,T^7]&=0, & [T^5,T^7]&=0, & [T^6,T^7]&=0, \notag \\
[T^1,T^5]&=0, & [T^4,T^2]&=0, & [T^2,T^6]&=0, & [T^5,T^3]&=0, & [T^3,T^4]&=0, & [T^6,T^1]&=0, \notag \\
[T^1,T^7]&=0, & [T^4,T^8]&=0, & [T^2,T^7]&=0, & [T^5,T^8]&=0, & [T^3,T^7]&=0, & [T^6,T^8]&=0.
\label{eq:R8_additional_ADHM-eq_II}
\end{align}

\end{appendix}

\end{document}